\documentclass[usenatbib, english]{mn2e}
\pdfoutput=1
\usepackage[fleqn]{amsmath}
\usepackage{array,multirow,graphicx}
\usepackage{txfonts}
\usepackage{lmodern}
\bibpunct{(}{)}{;}{a}{}{,}
\usepackage{floatrow}
\usepackage{array}
\usepackage{relsize}
\usepackage[font=small,labelfont=bf]{caption}
\usepackage{bm}

\usepackage{listings}
\newfloatcommand{capbtabbox}{table}[][\FBwidth]

\def\app#1#2{%
\mathrel{%
\setbox0=\hbox{$#1\sim$}%
\setbox2=\hbox{%
\rlap{\hbox{$#1\propto$}}%
\lower1.1\ht0\box0%
}%
\raise0.25\ht2\box2%
}%
}
\def\approxprop{\mathpalette\app\relax}

\title[Where do galaxies end?]{Where do galaxies end?\\  Comparing measurement techniques of hydrodynamic-simulation galaxies' integrated properties}
\author[A. R. H. Stevens et al.]{Adam R. H. Stevens$^1$\thanks{E-mail: astevens@swin.edu.au}, Marie Martig$^{1,2}$, Darren J. Croton$^1$, Yu Feng$^3$\\
$^1$Centre for Astrophysics \& Supercomputing, Swinburne University of Technology, Hawthorn, VIC 3122, Australia\\
$^2$Max-Planck-Institut f\"{u}r Astronomie, 69177 Heidelberg, Germany\\
$^3$McWilliams Center for Cosmology, Department of Physics, Carnegie Mellon University, Pittsburgh, PA 150213, USA}

\begin{document}

\pagerange{\pageref{firstpage}--\pageref{lastpage}} \pubyear{2014}

\maketitle

\label{firstpage}

\begin{abstract}
Using the suite of high-resolution zoom re-simulations of individual haloes by Martig et al., and the large-scale simulation \emph{MassiveBlack-II}, we examine the differences in measured galaxy properties from techniques with various aperture definitions of where galaxies end.  We perform techniques popular in the literature and present a new technique of our own, where the aperture radius is based on the baryonic mass profiles of simulated (sub)haloes.  For the average Milky-Way-mass system, we find the two most popular techniques in the literature return differences of order 30 per cent for stellar mass, a factor of 3 for gas mass, 40 per cent for star formation rate, and factors of several for gas accretion and ejection rates.  Individual cases can show variations greater than this, with the severity dependent on the concentration of a given system.  The average difference in integrated properties for a more general galaxy population are not as striking, but are still significant for stellar and gas mass, especially for optical-limit apertures.  The large differences that can occur are problematic for comparing results from various publications. We stress the importance of both defining and justifying a technique choice and discourage using popular apertures that use an exact fraction of the virial radius, due to the unignorable variation in galaxy-to-(sub)halo size.  Finally, we note that technique choice does not greatly affect simulated galaxies from lying within the scatter of observed scaling relations, but it can alter the derived best-fit slope for the Kennicutt-Schmidt relation.
\end{abstract}


\section{Introduction}
\label{sec:intro}
From humble beginnings \citep{carlberg90,katz92,evrard94}, hydrodynamic supercomputer simulations of the formation and evolution of galaxies and structure in the Universe have grown to be highly sophisticated, both in terms of their modelled physics and technical specifications \citep[e.g.][]{gimic,dimatteo12,sijacki12}. While the current state of the art makes such simulations an excellent tool for studying galaxy evolution and cosmology, interpreting the vast volumes of particle data output to conduct science presents many notable challenges.

To investigate the gross evolution of galaxies, it is informative to study their integrated properties.  Many surveys have been conducted with this as a central purpose, e.g. SDSS \citep{sdss} and 6dF \citep{6df}.  While there are established means of using simulations to predict observables \citep[e.g.][]{sunrise}, ultimately, observations attempt to measure quantities which should be `directly' measurable from simulations.  

In order to measure the integrated properties of a hydrodynamic-simulation galaxy, one needs a way of defining which particles/cells actually belong to the galaxy.   This sounds like a rather trivial task, but it is not. There is no solitary `right' way to define, for instance, `the size of the galaxy' or, rather, the boundary between the galaxy and the rest of the (sub)halo, nor is there to identify satellites or substructure that should not (yet) be considered part of the galaxy of interest.  For cosmological simulations that include many objects, one must also first face the task of structure location and halo definition \citep[see][]{knebe13b}.  Each decision carries an inevitable level of arbitrariness, partly driven by an incomplete definition of what a real galaxy is \citep[see][who propose a vote on the definition]{forbes11}.

An array of techniques for defining galaxies in hydrodynamic simulations has been used in the literature.  Some rely primarily on the results of subhalo finders \citep[e.g.][]{keres05,keres09,keres12,saro10,governato12,neistein12,sales12,haas13,moster13,munshi13}, which invoke overdensity, friends-of-friends (FoF), and/or gravitational binding criteria. Usually, this is coupled with restrictions on gas properties. It has also often been assumed that a galaxy will fall within some prescribed spherical aperture: examples include the radius at which the modelled surface brightness reaches some threshold \citep[e.g.][]{brook12,martig12}, an aperture of a fixed size in physical or comoving coordinates \citep[e.g.][respectively]{martig09,keres12}, or with a radius set to a fixed fraction of the virial radius \citep[e.g.][]{hirschmann12,sales12,scannapieco12,benitez13,marinacci13,roskar13}.

We aim to address the arbitrary choices involved in determining what constitutes a simulated galaxy and to compare results of integrated properties produced from an assortment of techniques, many of which have been used in publications. The goal is to provide a discussion on the motivations behind each technique, while addressing details of their implementation and determining which properties are most sensitive to technique choice, thereby realising the level of uncertainty associated with these measurements.  This tangentially builds on previous suggestions by \citet{guedes11} and \citet{munshi13} that stellar mass can vary by $\sim$20 per cent from summing the mass of all star particles within a simulated halo versus undertaking mock observations \citep[also see][who compare direct measurements with spectral energy distribution fits for particles within the same optically motivated radius]{obreja14}. Further, we aim to address the influence the techniques have on galaxy scaling relations.  We cover two different, but complementary, simulation regimes, considering both a suite of high-resolution zoom re-simulations of individual haloes \citep[][hereafter referred to as the M12 simulations]{martig12}, and subhaloes identified in the aptly massive cosmological simulation \emph{MassiveBlack-II} \citep{khandai14}.

In this paper, Section \ref{sec:sims} describes the simulations we have used to perform our analysis. Section \ref{sec:techs} outlines, in more detail, how galaxy particles are typically identified in hydrodynamic simulations.  Here, we cover specific techniques used in the literature, describing those we have used for our analysis, while outlining a new technique of our own.  We also describe how well those techniques succeed at capturing galaxies.  We provide comparative results of integrated property measurements from our analysis in Section \ref{sec:results}. We offer a brief discussion on what to make of the results in Section \ref{sec:discuss}. Section \ref{sec:conc} concludes the paper with a summary.

Following the common convention, throughout this paper, the virial radius of a halo is taken to be the radius at which the average internal density is $200 \rho_{\mathrm{crit}}$, where $\rho_{\mathrm{crit}}$ is the critical density of the Universe. This radius is denoted as $r_{200}$.  Except where the expansion factor $a = (1+z)^{-1}$ is explicitly written, all distances quoted in this paper are in physical coordinates, not comoving.  Because $h$ is a dimensionless constant and not a unit \citep[see][]{croton13}, we dissociate explicit factors of $h$ from relevant units, in favour of it either being attached to the \emph{number} in question instead or having its value substituted, whichever is more exact.

\section{The Simulations}
\label{sec:sims}
\subsection{The M12 simulations}
\label{ssec:zoomsims}
The M12 simulations are a set of 33 high-resolution zoom re-simulations run by \citet{martig12}, which each focus on a central, Milky-Way-size galaxy and its surrounding satellites.  Here, merger and diffuse-accretion histories were first extracted for each object of interest from a cosmological, dark-matter-only simulation. These histories were then re-simulated at higher resolution, where diffuse accretion was modelled assuming gas follows dark matter in an amount given by the cosmic baryon fraction.  Merging haloes were tracked in the simulation and replaced with higher-resolution haloes housing galaxies (made of a stellar disc and bulge, and a gas disc) upon their entrance of one of the re-simulated volumes.

The parent simulation was run using the code \textsc{ramses} \citep{ramses} in a periodic box of length $20h^{-1}a$ Mpc, with $512^3$ particles of mass $6.9 \times 10^6$ M$_{\odot}$, assuming $\Lambda$CDM parameters $\Omega_M = 0.3$, $\Omega_{\Lambda} = 0.7$, $h=0.7$, and $\sigma_8 = 0.9$.

The re-simulations were performed using a Particle-Mesh code, where gas dynamics was modelled with a sticky-particle scheme. Star formation followed a \citet{schmidt59} law, and supernova feedback and stellar mass loss were included (but not active galactic nucleus feedback). Particle masses are $3 \times 10^5$ M$_{\odot}$ for dark matter and $1.5 \times 10^4$ M$_{\odot}$ for gas and stars.\footnote{True for star particles formed during the simulations. Star particles existent from the initial conditions have mass $7.5 \times 10^4$ M$_{\odot}$.} Details on the zoom technique are outlined in \citet{martig09}. Each re-simulated box had a length of 800 kpc and spatial resolution of 150 pc.  The re-simulations started at $z=5$ and ran to $z=0$. For $z \lesssim 2$, the systems are sufficiently evolved such that measurements of the galaxies' properties are trustworthy.

We used a sample of 16 re-simulated galaxies, providing a fair representation of the diverse histories these simulations showcase spiral galaxies to have.  At $z=0$, the sample covers halo masses from 2.7 to $20 \times 10^{11}$ M$_{\odot}$ and bulge-to-total ratios between 0.02 and 0.53.  We analysed snapshots at 375-Myr intervals.

Gas particle densities were recovered using a refinable mesh, where particles within each cell were assigned the average density of that cell.  Temperatures were subsequently calculated using a polytropic equation of state \citep{teyssier10}:
\begin{equation}
\frac{T}{10^4\ \mathrm{K}} = 
\left\{
\begin{array}{rr}
\left( \frac{n_{\mathrm{H}}}{0.3\ \mathrm{cm}^{-3}} \right)^{-1/2}, & n_{\mathrm{H}} > 0.3\ \mathrm{cm}^{-3}\\
\multicolumn{2}{r}{1,\ \ \  n_{\mathrm{H}} \in [10^{-3},0.3]\ \mathrm{cm}^{-3}}\\
400\left( \frac{10^3\ n_{\mathrm{H}}}{\mathrm{cm}^{-3}} \right)^{2/3}, & n_{\mathrm{H}} < 10^{-3}\ \mathrm{cm}^{-3}\\
\end{array}
\right. ,
\label{eq:eos}
\end{equation}
\noindent where $n_{\mathrm{H}}$ is the number density of hydrogen atoms, which we approximated gas to be entirely composed of.  Given the piecewise nature of Equation~\ref{eq:eos}, throughout this paper, all gas particles with $T \leq 10^4$ K are considered `cold' for these simulations, with the remainder considered `hot'.

\subsection{MassiveBlack-II}
\label{ssec:mbii}
\emph{MassiveBlack-II} \citep{khandai14} is a cosmological, hydrodynamic simulation run with $2 \times 1792^3 \simeq 11.5$ billion particles in a $100 h^{-1}a$ Mpc periodic box, using the TreePM-SPH code \textsc{p-gadget} \citep[for a description of the earlier \textsc{gadget-2}, see][]{gadget2}.   The cosmology of the simulation followed $\Lambda$CDM with $\Omega_M = 0.275$, $\Omega_{\Lambda} = 0.725$, $h=0.702$, and $\sigma_8 = 0.8$.  Gas and dark matter particles have initial\footnote{Stars particles are created with mass $1.58 \times 10^6$ M$_{\odot}$, meaning a gas particle's mass is halved upon creating a star particle.} masses equal to $3.16 \times 10^6$ and $1.57 \times 10^7$ M$_{\odot}$, respectively, while gravitational softening occurred at $1.85 h^{-1}a$ kpc.  \emph{MassiveBlack-II} ran to $z=0$ and was a follow-up simulation to \emph{MassiveBlack} \citep{dimatteo12} which ran to $z=4.75$ using 8 times as many particles, at lower mass resolution, in a volume 151 times larger.  Star formation and stellar feedback followed \citet{springelhernquist}.  A full consideration of black holes and active galactic nucleus feedback was also included  \citep[following][]{dimatteo05,springel05,dimatteo08}.  

We analyse one snapshot of \emph{MassiveBlack-II} at $z=0.0625$, measuring results from subhaloes whose gas (hot + cold) and stellar masses are each between $10^8$ and $10^{12}$ M$_{\odot}$.  This ensured that each subhalo assessed was well resolved (by at least 63 gas and star particles each).  This gave a sample of 224,585 galaxies.

To maintain consistent definitions of `cold' and `hot' throughout this work, densities of gas particles in \emph{MassiveBlack-II} were calculated via the same means as in the M12 simulations, where temperatures were subsequently calculated with Equation \ref{eq:eos}.\footnote{A proper treatment of hydrogen number density could affect the number of hot and cold particles for \emph{MassiveBlack-II} here, where the difference would be negligible for the M12 simulations.}

\section{Techniques for Measuring Galaxy Properties}
\label{sec:techs}
\begin{table*}[t!]
\centering
\begin{tabular}{l c r l l r r l}\hline
Ref. & Type & Best mass res. & Subhalo finder & Aperture & $T$ max. & $n_{\mathrm{H}}$ or $\rho$ min. & Properties \\
& & $\times 10^3$ M$_{\odot}$ & or observing code & radius &  &  & \\\hline
$[1]$ & P & $1,850.0 h^{-1}$ & FoF & --------- & --------- & 0.1 cm$^{-3}$ & SM, GM\\
$[2]$ & Z &  $170,000.0 h^{-1}$ & \textsc{subfind} $[22]$ & --------- & --------- & --------- & SM, SFR\\
$[3]$ & Z & $1,400.0 h^{-1}$ & \textsc{subfind} & $0.15 r_{200}$ & --------- & --------- & SM, GM\\
$[4]$ & Z &  $\sim$100.0\ \ \ \ \ \ & \textsc{subfind} & --------- & $2 \times 10^4$ K & --------- & SM, GM\\
$[5]$ & P & $86,600.0h^{-1}$ & \textsc{subfind} & --------- & $10^5$ K & 0.1 cm$^{-3}$ & SM, GM, SFR\\
$[6]$ & P & $740.0 h^{-1}$ & \textsc{subfind} & $30 h^{-1}a$ kpc & $3 \times 10^4$ K & $2 \times 10^3 \bar{\rho}_{\mathrm{bary}}$ & SM, GM, SFR\\
$[7,8]$ & Z & $0.4$\ \ \ \ \ \ & \textsc{ahf} $[23,24]$ & --------- & --------- & --------- & SM, GM\\
$[9,10]$ & P & $1,700.0$\ \ \ \ \ \ & \textsc{skid} $[25,26]$ & --------- & $3 \times 10^4$ K & $10^3 \bar{\rho}_{\mathrm{bary}}$ & SM, GM, SFR, GAR\\
$[11]$ & Z & $20.0$\ \ \ \ \ \ & \textsc{sunrise} $[27]$ & --------- & --------- & --------- & SM\\
$[12]$ & Z & $15.0$\ \ \ \ \ \ & \textsc{pegase} $[28]$ & $R_{25}$ $g$-band & --------- & --------- & SM, GM, SFR, GAR*\\
$[13]$ & Z & $4.8$\ \ \ \ \ \ & \textsc{sunrise} & $R_{25}$ $i$-band & --------- & --------- & SM, GM*\\
$[14]$ & Z & $75.0$\ \ \ \ \ \ & \textsc{grasil-3d} $[29]$ & $2R_P$ & --------- & --------- & SM, SFR\\
$[15]$ & Z & $89.0$\ \ \ \ \ \ & --------- & $0.1 r_{200}$ & --------- & --------- & SM, SFR*\\
$[16]$ & Z & $230.0$\ \ \ \ \ \ & --------- & $0.1 r_{200}$ & --------- & --------- & SM, GM, SFR \\
$[17]$ & Z & $10.0$\ \ \ \ \ \ & --------- & $0.1 r_{200}$ & $1.5 \times 10^4$ K & --------- & GM, SFR, GAR \\
$[18]$ & Z & $200.0$\ \ \ \ \ \ & --------- & $0.1 r_{200}$ & $10^5$ K & --------- & SM, GM, SFR\\
$[19]$ & Z & $4,200.0 h^{-1}$ & --------- & $0.1 r_{200}$ & \multicolumn{2}{c}{$\log{T} < 0.3\log{\rho} + 3.2$} & SM, GM, SFR\\
$[20]$ & Z & $60.5$\ \ \ \ \ \ & --------- & $0.15 r_{200}$ & --------- & --------- & SM, SFR\\
$[21]$ & Z & $21.0$\ \ \ \ \ \ & --------- & $25$ kpc & --------- & --------- & SM, GM, SFR\\\hline
\end{tabular}
\caption{Summary of techniques used for measuring integrated properties of galaxies published in the literature. `Type' represents whether the study was of a periodic-box (P) or zoom (Z) simulation.  The third column provides the best resolution of baryonic mass used in each reference. The quoted resolution applies to the results presented in the papers, where some conduct resolution studies that involve higher resolutions.  An aperture radius of $R_{25}$ is the radius at which the modelled surface brightness reaches 25 mag arcsec$^{-2}$ in the listed band.  $R_P$ denotes the Petrosian radius \citep{blanton01}.  The sixth and seventh columns provide the maximum allowable temperature and minimum allowable density, respectively, for gas particles. $\bar{\rho}_{\mathrm{bary}}$ is the mean baryonic density.  Column eight lists the properties measured in the papers, including stellar mass (SM), gas mass (GM), star formation rate (SFR), and gas accretion rate (GAR).  Properties with a star were measured via a different method to the one in this table.  Where entries are not applicable, a horizontal line has been placed. References are as follows:\newline [1] \citet{haas13}; [2] \citet{saro10}; [3] \citet{sales12}; [4] \citet*{moster13}; [5] \citet{neistein12}; [6] \citet{keres12}; [7] \citet{governato12}; [8] \citet{munshi13}; [9] \citet{keres05}; [10] \citet{keres09}; [11] \citet{guedes11}; [12] \citet{martig12}; [13] \citet{brook12}; [14] \citet{obreja14}; [15] \citet{roskar13}; [16] \citet*{marinacci13}; [17] \citet{dekel13}; [18] \citet{scannapieco12}; [19] \citet{hirschmann12}; [20] \citet{benitez13}; [21] \citet{martig09}; [22] \citet{subfind}; [23] \citet{ahf1}; [24] \citet{ahf2}; [25] \citet{skid1}; [26] \citet{skid2}; [27] \citet{sunrise}; [28] \citet{pegase}; [29] \citet{grasil}.}
\label{tab:techs}
\end{table*}

As galaxies occupy (sub)haloes, particles belonging to a galaxy should be a subset of the baryons belonging to a (sub)halo.  Baryons in (sub)haloes can be broken into three populations: those in the galaxy of interest, those in other galaxies within the same (sub)halo (\frenchspacing{i.e. satellites}), and those diffusely occupying the rest of the (sub)halo.  As has been done in the literature, this breakdown can be achieved with an automated, generally applicable method involving three steps: 

\begin{enumerate}
\item Use of a subhalo finder. Subhalo finders not only locate and provide the (sub)haloes that galaxies of interest occupy, but are also proficient at identifying satellite systems to be stripped.
\item Temperature and density restrictions on gas, as gas within a galaxy should be relatively dense and cool.
\item A cut with a spherical aperture.  This sets a boundary between the galaxy and the rest of the (sub)halo, classifying which of the remaining star and cold gas particles are part of the galaxy.  Spherical apertures are both simple in their implementation and a fair shape to broadly apply to all galaxies.
\end{enumerate}

There are many widely used, well-established subhalo-finding codes available \citep[for a comparative study on popular codes' abilities to locate galaxies, see][]{knebe13a}, while the cold/hot temperature boundary for gas is often motivated by how temperature was treated within the simulation of interest.  The choice of spherical aperture is hence the most contentious step.  This ultimately determines where the galaxies end: a choice that has shown great variation throughout the literature.  Indeed, the primary goal of this paper is to quantify the differences in integrated properties returned from these aperture choices.  We stress the value the above methodology has in being, in theory, blindly applicable; this is mandatory for analysing large simulation datasets efficiently.

\subsection{Aperture choices in the literature}

We present a summary of the techniques for defining particles/cells associated with simulated galaxies in the literature in Table \ref{tab:techs}.  Many of these cases include only one or two of the ideal steps listed above.  For instance, while most of the examples we list include the application of an aperture, it has also been popular to not apply one at all.  We note techniques beyond what we have listed exist in the literature as well \citep[e.g.][who decompose the galaxy into disc and bulge particles]{domenech12,few12}.  The aperture choices listed in Table \ref{tab:techs} can be broken into categories of an optical limit ($R_{25}$), fixed aperture (comoving or physically fixed), and fraction of the virial radius.  We discuss the potential motivations and short falls of each in turn below.

Optical limits describe where a galaxy would appear to end against a sky background, and hence are appropriate for more-direct comparisons to observations.  However, they do not aim to probe where a galaxy \emph{actually} ends.  While integrated properties derived from this technique certainly have value, they will not necessarily be the `true' integrated properties of a given galaxy, which are of interest here.

For a small sample of galaxies of similar shapes and masses at some epoch, it might be reasonable that each galaxy ends at an approximately equivalent radius.  However, using an aperture of fixed comoving size to define the end of galaxies does not work on a general basis, simply as galaxies come in a range of sizes.  This is at least preferable to an aperture of fixed physical size though, as it considers the average growth of galaxies with time.

Using a fraction of the virial radius is the most common aperture method in the literature.  For this to be appropriately motivated, there would need to be a strong correlation between the size of galaxies and their parent (sub)haloes.\footnote{None of the literature examples we present describe any motivation behind using a precise fraction of $r_{200}$, so we are left to speculate.}  A study by \citet{kravtsov13}, which compared the cumulative number density of observed galaxies in the local Universe as a function of their size to the same distribution of simulated dark matter haloes, found a trend between a galaxy's stellar half-mass radius, $r_{1/2}$, and its virial radius. Specifically, the author showed
\begin{equation}
\label{eq:rhalf}
4.5 < \frac{r_{1/2}}{r_{200}} \times 10^3 < 45\ . 
\end{equation}
\noindent This relationship is consistent with the model of \citet{mmw}, who suggested the radii of discs should be proportional to $\lambda r_{200}$, where $\lambda$ is the dimensionless spin parameter of the galaxy's parent halo. The scatter one expects from the probability distribution of $\lambda$ strongly correlates to the allowable range in Equation \ref{eq:rhalf} \citep{kravtsov13}.

This result has been interpreted by others \citep[e.g.][]{stringer13} to mean there is a tight correlation between the size of a galaxy and its parent (sub)halo.  We find several points to be troubling about this conclusion:

\begin{enumerate}
\item The relation only considers the stellar component of any galaxy.  Galaxies also consist of gas.  The distribution of gas should be included in the interpretation of where a galaxy ends.
\item There is a spread of an order of magnitude for the correlation.  To use it to infer the size of a galaxy hence carries a large uncertainty.
\item A correlation for a half-mass radius only infers a correlation to a full-mass radius if the baryonic density profiles of all galaxies are equivalent, which they are not (see below).
\end{enumerate}

The majority of examples in Table \ref{tab:techs} study a limited number of integrated properties, some of which are not measured with equivalent techniques.  We emphasize that, especially if one wishes to measure multiple properties that are derivatives of time (\frenchspacing{e.g. star formation rate, gas ejection rate, gas accretion rate, star death rate}), a solitary technique that is self-consistent in a mass-conserving sense should be used to measure all properties.

\subsection{Employed subhalo finders}
\label{ssec:gravtechs}

We employed \textsc{ahf}\footnote{\textsc{amiga} Halo Finder. Downloadable at http://popia.ft.uam.es/AMIGA/} \citep*{ahf1,ahf2} to identify substructure in the M12 simulations.  \textsc{ahf} operates by first finding density peaks in a simulation using a refinable-mesh scheme, then places spherical apertures around those regions based on an input overdensity criterion (typically $200\rho_{\mathrm{crit}}$).  Finally, the program performs an unbinding procedure on the particles, where only particles with kinetic energy (kinetic + thermal for gas) less than their potential energy relative to the overdensity are considered bound to the structure.  We ensured the refinement process of \textsc{ahf} could only go down to a cell length equal to the gravitational softening of the M12 simulations (150 pc).  \textsc{ahf} offers the choice of a critical velocity to determine particles as unbound; we chose this to be exactly equal to the escape velocity calculated within the program.  In accordance with the previously published techniques, all subhaloes have been stripped from the \textsc{ahf} data presented.  

Subhaloes in \emph{MassiveBlack-II} were found with \textsc{subfind} \citep{subfind}.  \textsc{subfind} identifies subhaloes within FoF groups by enclosing areas with isodensity contours that traverse saddle points, then groups particles whose energies are sufficiently low to be gravitationally bound (the same final step as \textsc{ahf}).  

While, to first order, \textsc{ahf} and \textsc{subfind} achieve the same goal, their methods of identifying substructure are not identical.  Comparison studies of these and other subhalo finders \citep{knebe11,knebe13a,onions12} have shown their results to be in good agreement, however.  As such, the impact of using them on the different simulations is small when compared to the much larger variations between the aperture techniques (see below).

The primary difference between the codes is that \textsc{ahf} uses spherical apertures to define the extent of subhaloes (for our results, this is $r_{200}$).  There is no change to our results from \emph{MassiveBlack-II} if an aperture cut at $r_{200}$ is additionally made to the \textsc{subfind} particles, however.  This is because the properties we assess are only concerned with particles that naturally exist toward the centre of subhaloes; for all results involving use of a subhalo finder, integrated gas masses and transfer rates only consider cold gas.  For properties concerning dark matter and hot gas, this additional aperture cut would have some effect.

\subsection{The baryonic-mass-profile (BaryMP) technique}
\label{ssec:barymp}

\begin{figure}[t!]
\centering
\includegraphics[width=\textwidth]{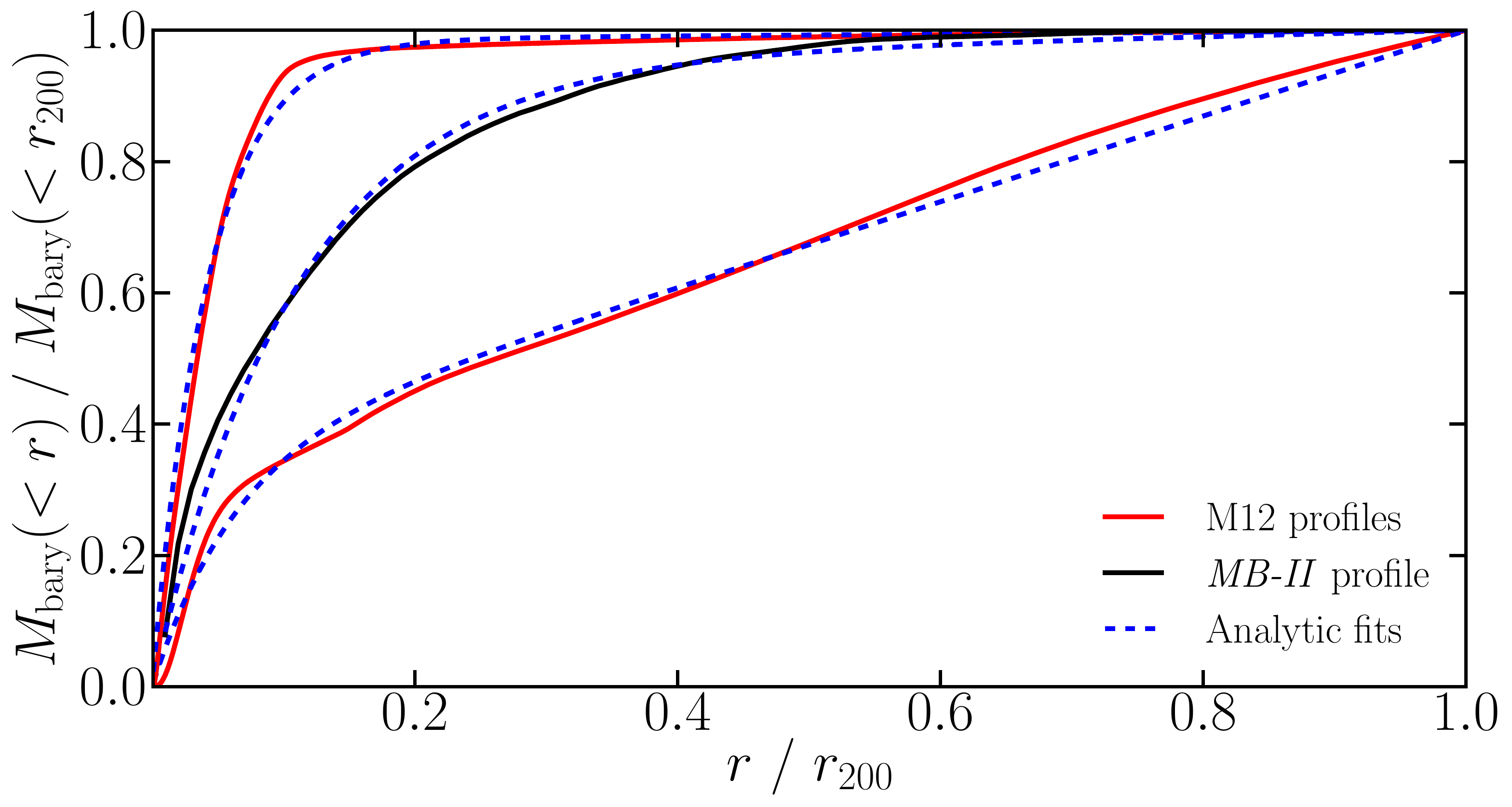}
\caption{Cumulative baryonic mass profiles for example (sub)haloes from the M12 and \emph{MassiveBlack-II} simulations (solid curves).  Substructure identified by the subhalo finders and hot gas were stripped prior to building these profiles.  Each profile can be approximated by an analytic function (dashed curves), given by Equation \ref{eq:barymp}.}
\label{fig:analprof}
\end{figure}

We propose a new technique for determining the size of the spherical aperture that defines the end of a galaxy.  Ideally, the end of a galaxy should be directly informed by the way in which baryons are distributed in the (sub)halo of interest.  Moreover, we see no reason to bias the size on the stellar distribution over the cold gas distribution or vice versa.  As such, we use the cumulative one-dimensional baryonic (stars + cold gas) mass profiles of each respective (sub)halo (without substructure) to infer their aperture sizes.

From both the M12 and \emph{MassiveBlack-II} galaxies, we discovered the baryonic mass profiles for all the simulated (sub)haloes follow a common form.  This can be analytically approximated as
\begin{equation}
\label{eq:barymp}
\frac{M_{\mathrm{bary}}(<r)}{M_{\mathrm{bary}}(<r_{200})} = (1 - e^{-br/r_{200}})(m \frac{r}{r_{200}} + k)\ ,
\end{equation}
\noindent where $b$ is defined to be positive, and $0 \leq m < 1$ (see below and Appendix \ref{app:algorithm}).  Because the left side of Equation \ref{eq:barymp} must be 1 when $r=r_{200}$, $b$ and $m$ are the only 2 parameters, where
\begin{equation}
\label{eq:c}
k = (1-e^{-b})^{-1} - m\ .
\end{equation}
\noindent We provide examples of baryonic mass profiles from the simulations placed against their analytic counterparts with best-fitting parameters in Fig. \ref{fig:analprof}.  For our sample of \emph{MassiveBlack-II} galaxies, the best-fitting $b$ parameters produce a smooth, approximately Gaussian distribution, with mean and standard deviation values of 14.2 and 3.8, respectively.  75 per cent of these systems also have $m<0.1$; the parameter is more important for the other 25 per cent of cases.

The $(m \frac{r}{r_{200}} + k)$ term in Equation \ref{eq:barymp} describes a straight-line asymptote that the function approaches as $r$ increases.   For large $r$ then, the cumulative baryonic mass effectively goes linearly with $r$ (with gradient $m$); in other words, baryonic density goes as $r^{-2}$.  This follows the relation for an isothermal sphere and also holds at the scale radius for dark matter haloes in virial equilibrium \citep{nfw}.  We define the diffuse halo component to occupy the region of the actual profiles where the gradient is roughly constant (for a profile void of substructure or abnormalities, this would equate to the asymptotic region of the analytic approximation).  By extension, all (cold) baryons internal to this radius constitute the galaxy.

With the above motivation, we have devised an algorithm to determine the radius where the gradient of the baryonic mass profiles of (sub)haloes is sufficiently close to constant, $r_{\mathrm{BaryMP}}$. This is done via iterative straight-line fits to the outer parts of the profiles (Equation \ref{eq:barymp} is not explicity used).   We provide a thorough description and code for the BaryMP algorithm in Appendix \ref{app:algorithm}.  We have subsequently employed this aperture technique for measuring the integrated properties of simulated galaxies.

\begin{figure}[t!]
\centering
\includegraphics[width=\textwidth]{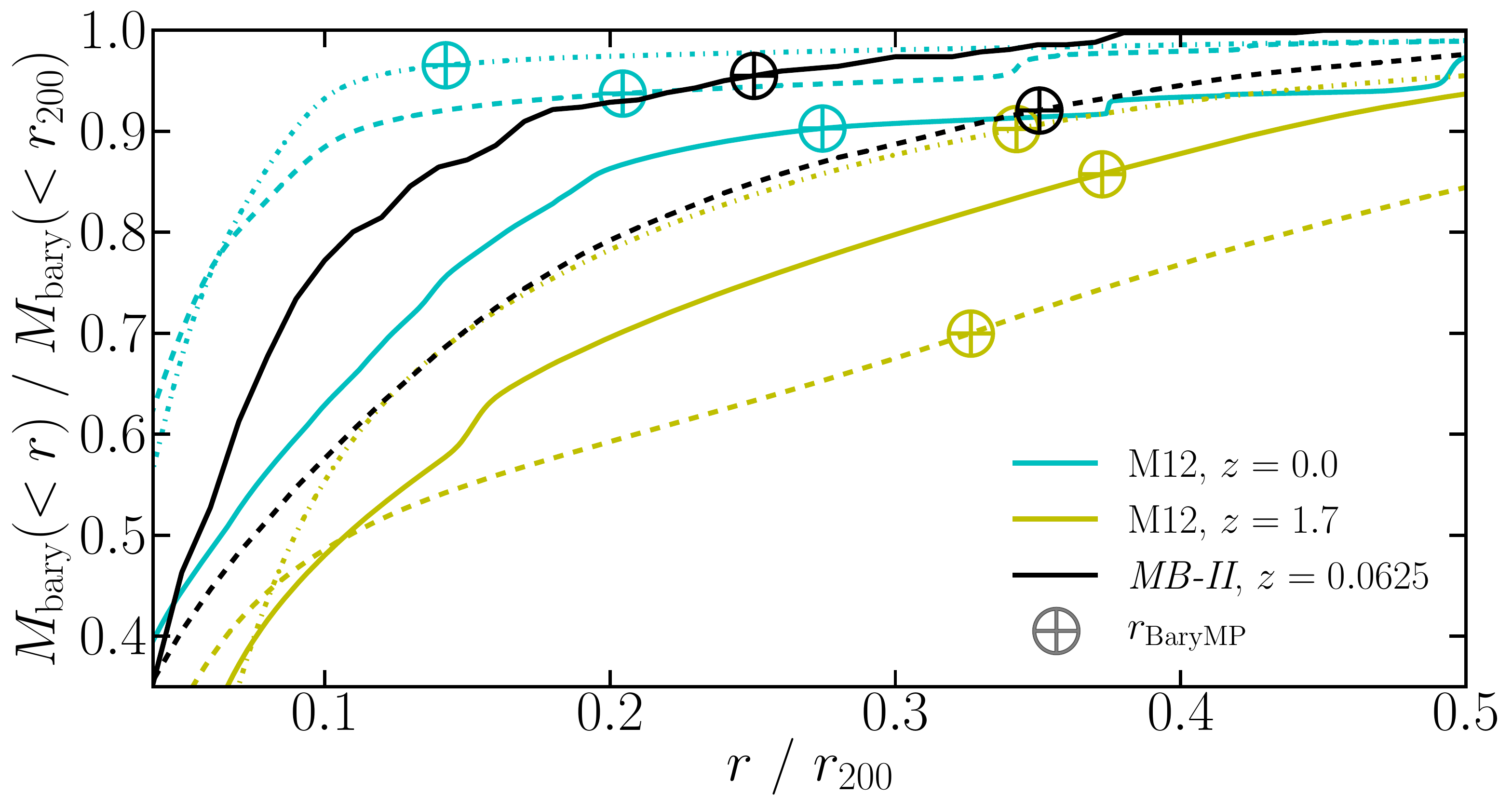}
\caption{Examples of baryonic mass profiles for \emph{MassiveBlack-II} subhaloes and M12 haloes, the latter at two epochs.  Solid, dashed, and dot-dashed lines are used to make the profiles differentiable.  The plus-filled circle on each profile shows where the aperture radius is determined from the BaryMP algorithm.}
\label{fig:propvrad}
\end{figure}

Fig. \ref{fig:propvrad} shows where the BaryMP aperture cuts are taken for example profiles from \emph{MassiveBlack-II} and two epochs of the M12 simulations.  We note that, for these examples, the aperture radii returned are various fractions of their respective virial radii.  At $z \sim 0$, we typically find $0.2 \lesssim r_{\mathrm{BaryMP}} / r_{200} \lesssim 0.4$, with only one occasion from M12 in agreement with an aperture of $0.15r_{200}$.  At higher redshift for M12, aperture radii consistently exceed $0.3r_{200}$.

\subsection{Employed aperture techniques}
\label{ssec:aperture}
In this paper, we closely mimic many of the techniques presented in Table \ref{tab:techs} \citep*[namely from][]{brook12,governato12,keres12,sales12,benitez13,munshi13,roskar13,marinacci13}. The primary aim of this is to indicate how the relevant publications' results should be compared with each other and our BaryMP technique.  We summarise the aperture techniques we apply in Table \ref{tab:apertures}.  We note that to keep the number of test techniques limited and to assess the most ideal cases, wherever subhalo finders are used, hot gas is also stripped.

\begin{table}[t!]
	\centering
	\begin{tabular}{l c c}\hline
		Aperture radius & M12 & \emph{MB-II}\\\hline
		$0.1r_{200}$ & D & S\\
		$0.15r_{200}$ & D, S & S\\
		$R_{i25}$ & D & S\\
		$30h^{-1}a$ kpc & S & S\\
		BaryMP & S & S\\\hline
	\end{tabular}
	\caption{Aperture techniques applied for our analysis, noting for each simulation set (M12 and \emph{MassiveBlack-II}) which have been used as direct aperture techniques (denoted by D) and which have been used in combination with a subhalo finder and gas temperature restriction (denoted by S).  $R_{i25}$ is the radius at which the surface brightness profile reaches $25$ mag arcsec$^{-2}$ in the $i$-band.}
\label{tab:apertures}
\end{table}

We refer to techniques that only included an aperture cut from the full simulation data as `direct aperture techniques'. These involve no attempt to remove satellites that fall in the aperture, nor is there a requirement to separate gas by temperature or density.  Direct aperture techniques were employed in some of the literature examples presented in Table \ref{tab:techs} \citep*{benitez13,marinacci13,roskar13}.

To study galaxies in a cosmological simulation like \emph{MassiveBlack-II}, a subhalo finder is needed to locate galaxies initially, regardless of whether there is intent to apply direct aperture techniques. In addition to the fact that no direct aperture techniques were performed on periodic-box simulations listed in Table \ref{tab:techs}, we have only applied them to the M12 simulations. Table \ref{tab:apertures} lists the instances where direct apertures have been applied.

For the direct aperture techniques we employed that use a fraction of the virial radius, $r_{200}$ values were calculated considering all particles, again irrespective of whether they were part of any substructure. $r_{200}$ values were recalculated without the substructure when subhalo finders were used.  Typically, the difference in these values is negligible.

For an optical radius, we find $R_{25}$ values in the $i$-band (hereafter $R_{i25}$).  To calculate these radii, we used the simple stellar population evolution model of \citet{maraston05}\footnote{Downloadable at http://www-astro.physics.ox.ac.uk/ $\sim$maraston/Claudia\%27s\_Stellar\_Population\_Models.html} with a \citet{kroupa01} initial mass function to produce spectral energy distributions for each star particle.  We subsequently applied the $i$-band filter and integrated the spectra to produce $i$-band luminosities for each particle.  Solar metallicity was assumed for the M12 star particles, while metallicity was tracked in \emph{MassiveBlack-II}.  Surface brightness profiles used to evaluate $R_{i25}$ were built using the AB magnitude system irrespective of redshift, \frenchspacing{i.e. assuming} galaxies were being viewed face-on (defined below) at a near enough distance where angular-diameter and luminosity distances are negligibly different.\footnote{In this regime, (apparent) surface brightness has no dependence on distance.  This can also be thought of as building ``absolute'' surface brightness profiles, treating high-redshift systems fairly.}  No additional dust modelling was included.

\begin{figure*}[t!]
\centering
\includegraphics[width=0.49\textwidth]{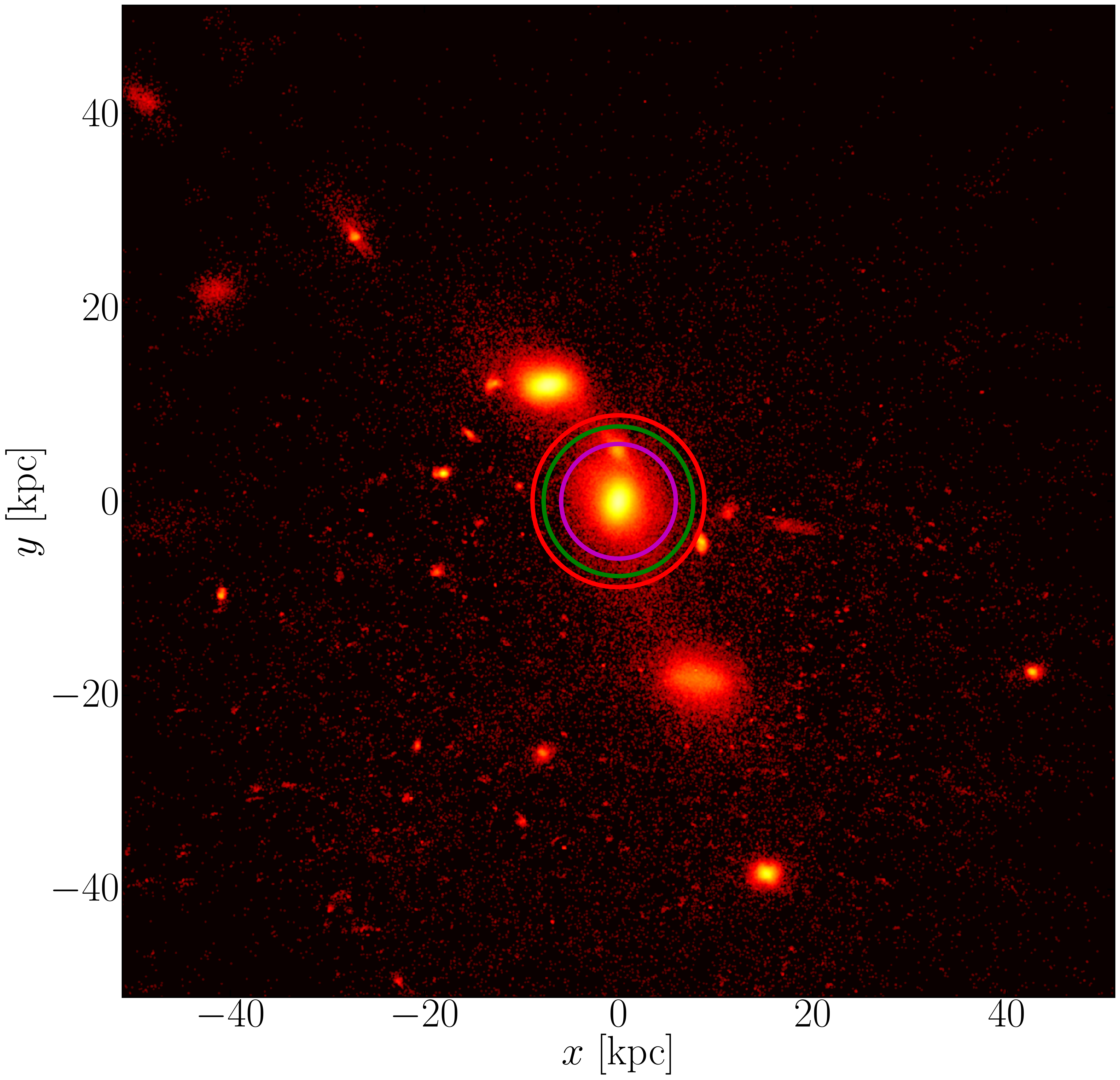}
\includegraphics[width=0.49\textwidth]{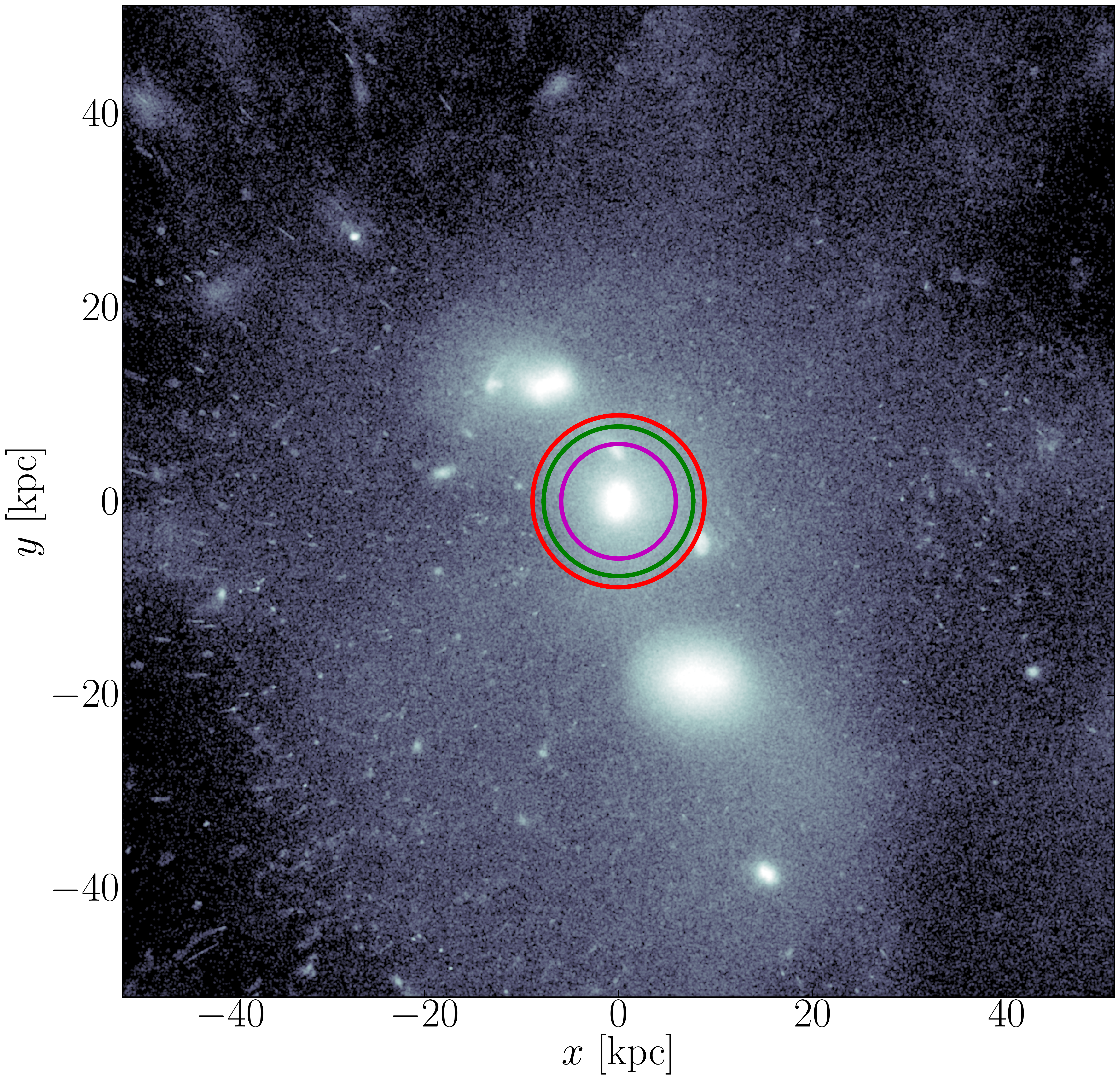}
\includegraphics[width=0.49\textwidth]{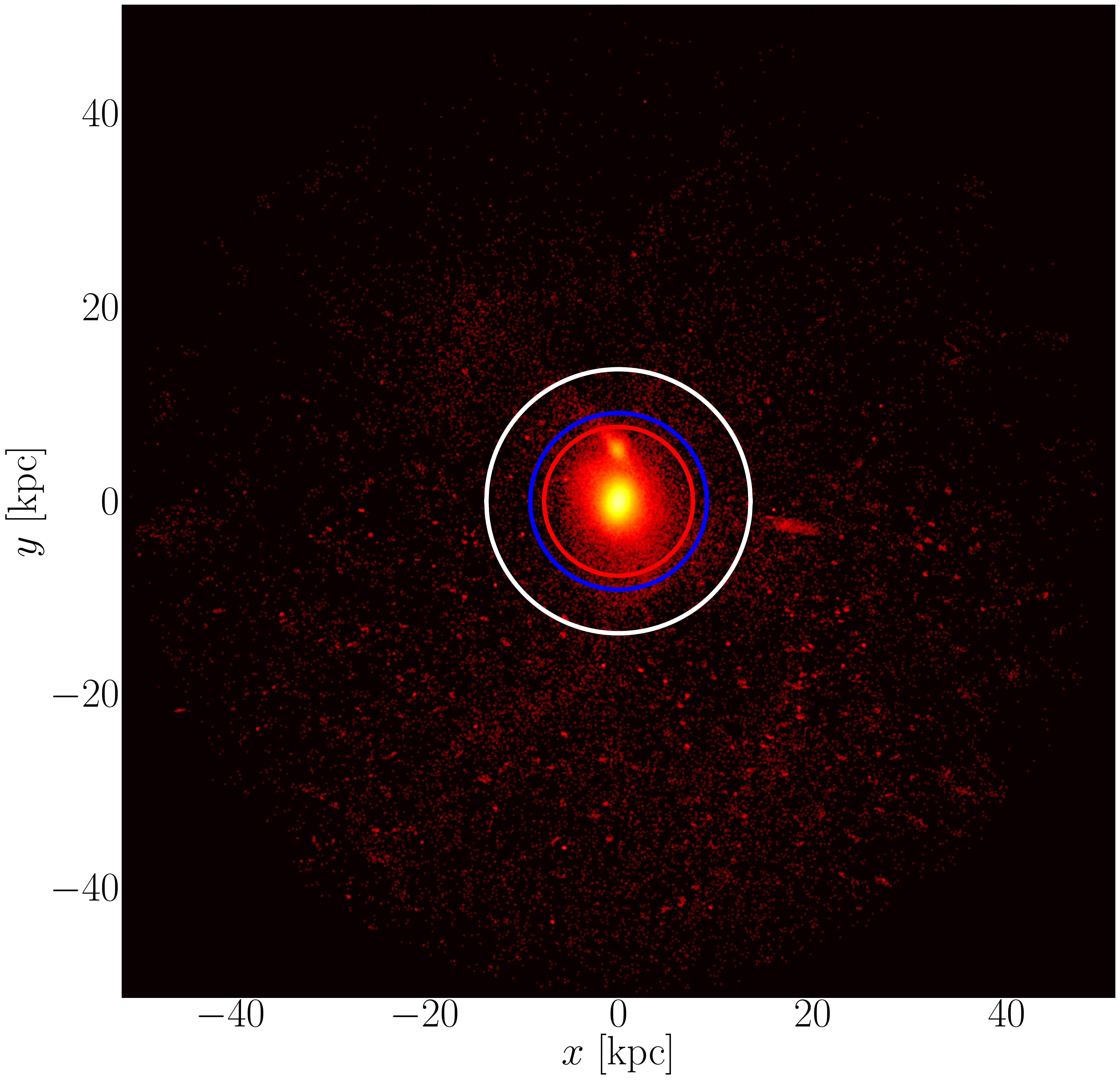}
\includegraphics[width=0.49\textwidth]{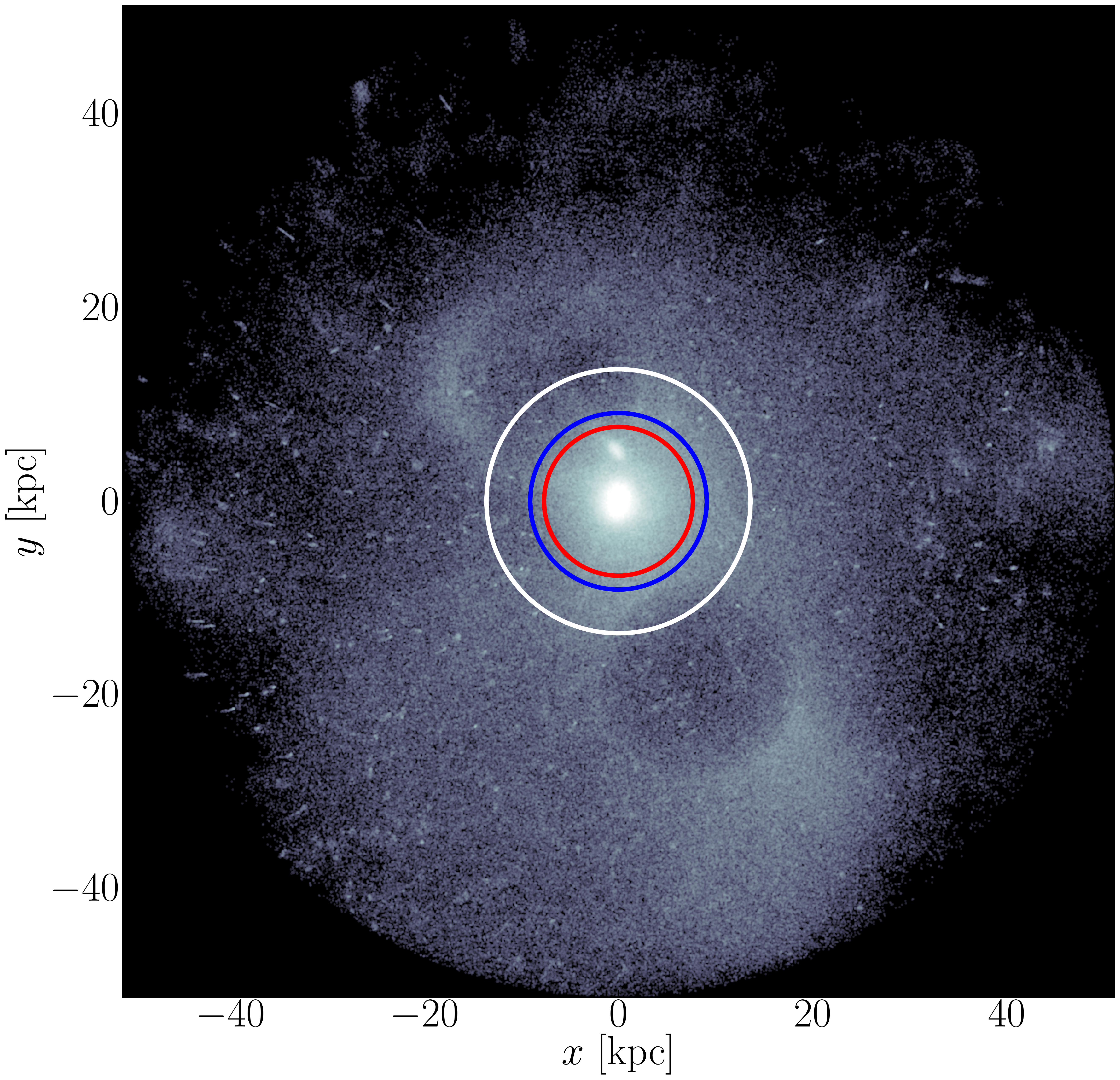}
\caption{Illustrations of one of the M12 simulations at $z = 2.15$. Left panels display stars, with black pixels indicating a column density $\leq 10^{-1}$ M$_{\odot}$ pc$^{-2}$ and white pixels $\geq 10^{3.5}$ M$_{\odot}$ pc$^{-2}$. Right panels display gas from $10^{-1}$ to $10^2$ M$_{\odot}$ pc$^{-2}$. The width, height, and depth of each frame is equal to $2r_{200}$, with $r_{200}$ calculated after stripping substructure and hot gas. Top panels display all particles within the image frame. Bottom panels only show particles associated with the main halo according to \textsc{ahf} (substructure and hot gas stripped). Apertures have been overdrawn in the following colour code (in order of smallest to largest): magenta = $0.1r_{200}$, green = $R_{i25}$, red = $0.15r_{200}$, blue = BaryMP, white = $30h^{-1}a$ kpc.}
\label{fig:images1}
\end{figure*}

\begin{figure*}[t!]
\centering
\includegraphics[width=0.49\textwidth]{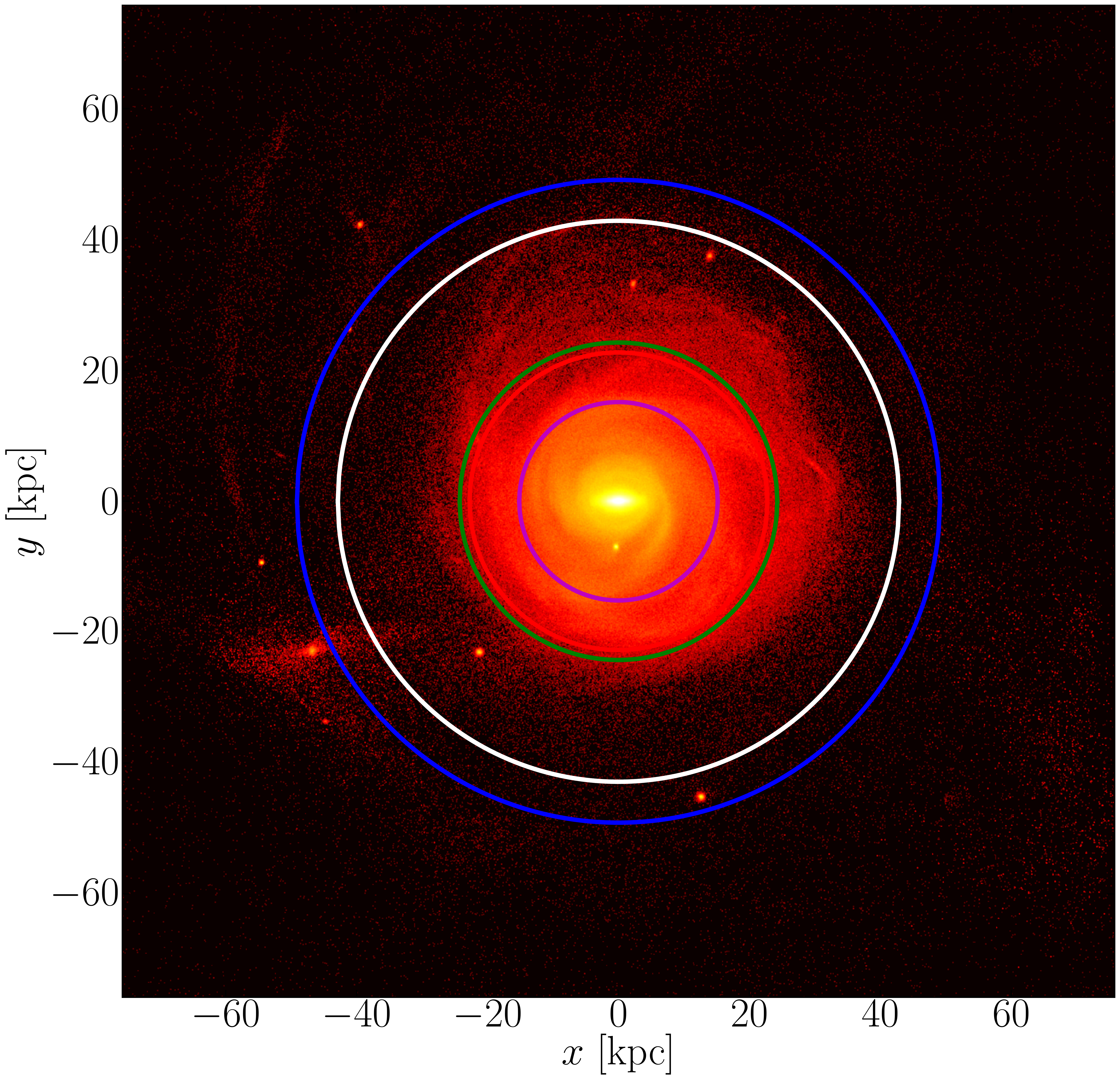}
\includegraphics[width=0.49\textwidth]{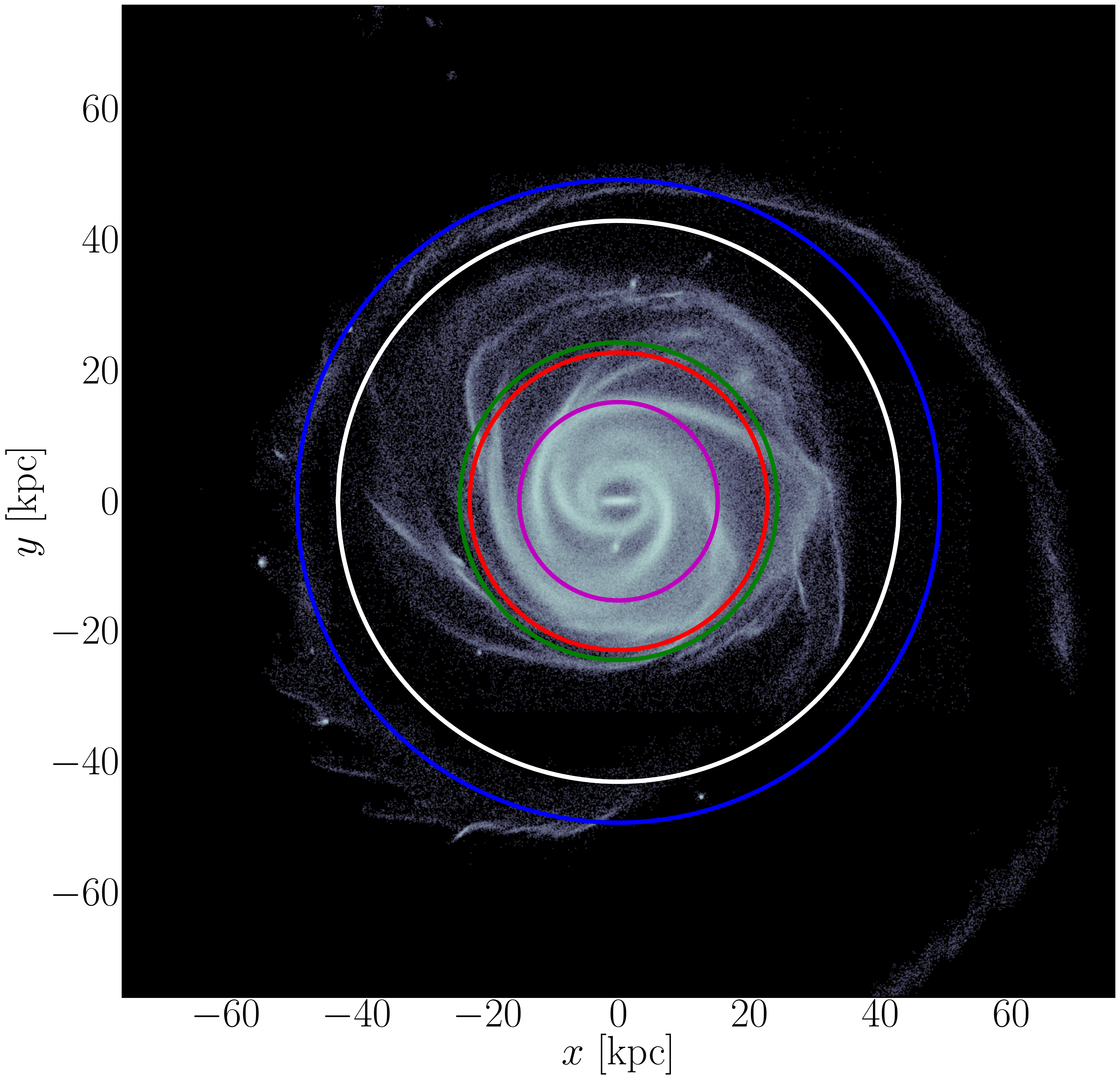}
\caption{The same M12 simulation as in Fig. \ref{fig:images1} but at $z=0$, with stars on the left and gas the right.  All \textsc{ahf}-identified substructure and hot gas has been stripped from these images (despite what appear to be satellite systems in the frame).  Each frame has width, height, and depth of $r_{200}$.  We exclude images prior to stripping substructure and hot gas, as, visually, there is little difference for this case.  Colour code for apertures is: magenta = $0.1r_{200}$, red = $0.15r_{200}$, green = $R_{i25}$, white = $30h^{-1}a$ kpc, blue = BaryMP.  Intensity scale matches Fig. \ref{fig:images1}.}
\label{fig:images2}
\end{figure*}

\begin{figure*}[t!]
\centering
\includegraphics[width=0.49\textwidth]{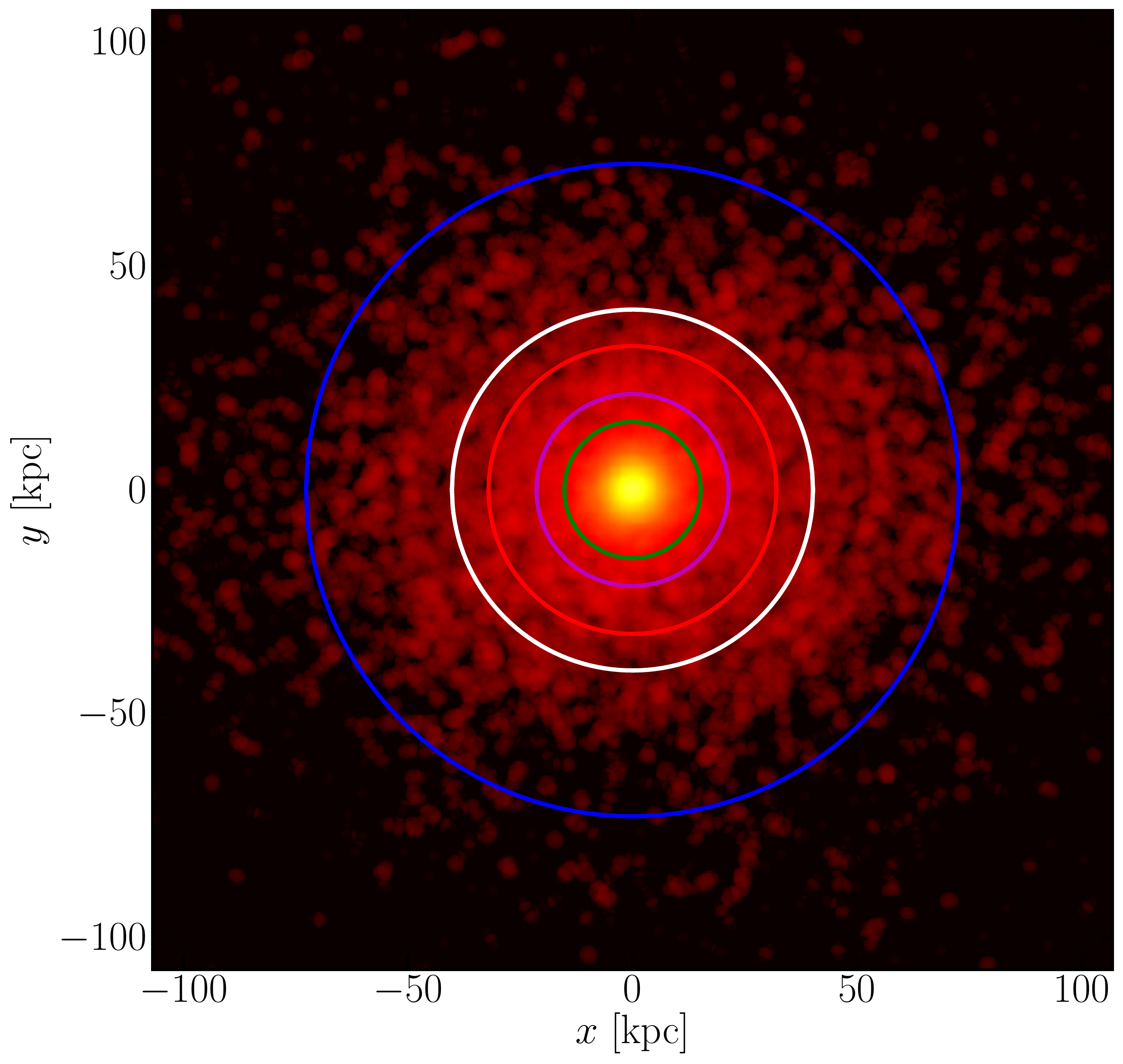}
\includegraphics[width=0.49\textwidth]{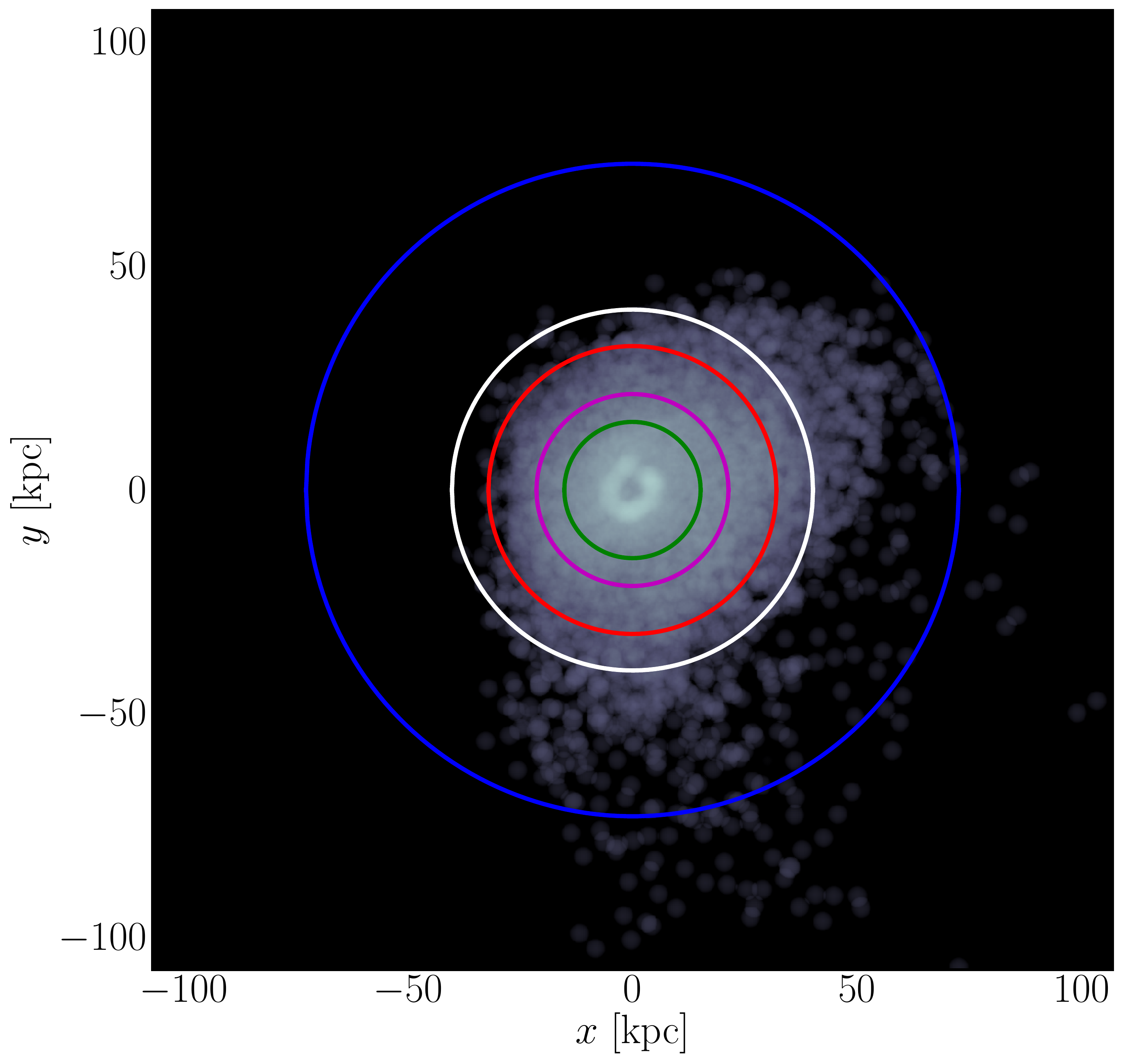}
\caption{Stars (left) and cold gas (right) of an example subhalo from \emph{MassiveBlack-II} at $z=0.0625$ with integrated masses from each particle species comparable to those in the M12 simulations.  Frame width, height, and depth are each $r_{200}$.  Intensity scale is identical to Figs. \ref{fig:images1} and \ref{fig:images2}.  Colour code for apertures is: green = $R_{i25}$, magenta = $0.1r_{200}$, red = $0.15r_{200}$, white = $30h^{-1}a$ kpc, blue = BaryMP.}
\label{fig:images3}
\end{figure*}

Figs. \ref{fig:images1} and \ref{fig:images2} present example images of an M12 simulation at two epochs with apertures from Table \ref{tab:apertures} overdrawn.  Fig. \ref{fig:images3} similarly presents an example from \emph{MassiveBlack-II}.  All three of these show `face-on' images, \frenchspacing{i.e. where} the net angular momentum vector of the substructure-stripped star particles within $r_{200}$ of the (sub)halo points out of the page toward the reader.  Fig. \ref{fig:images1} also highlights the potentially significant effect of removing substructure identified by \textsc{ahf}.

Fig. \ref{fig:images1} presents an example where, by eye, there is a distinguishable end to the stellar component of the galaxy, while there appears to be no end to the cold gas, which fills the entire halo.  With temperature restrictions rendered unhelpful to resolve this (and, by extension, density restrictions, due to the polytropic equation of state), there is no distinction between galaxy and halo without the use of an aperture.  Arguably, many of the apertures presented do a reasonable job of encompassing the stellar component of the galaxy, but it is not obvious from imaging alone which best encompasses the cold gas.  As such, the aperture technique applied needs to be solidly motivated.

Figs. \ref{fig:images2} and \ref{fig:images3} show how several of the aperture techniques can exclude a significant portion of the galaxy at lower redshift (as quantified in Section \ref{ssec:relm12}).  Specifically, $0.1r_{200}$ and $0.15r_{200}$ consistently underestimate the extent of galaxies.  Despite being the most observationally motivated technique, $R_{i25}$ also frequently excludes parts of the galaxy.  While Figs. \ref{fig:images1} and \ref{fig:images2} provide examples where a cut at $30h^{-1}a$ kpc seems reasonable, Fig. \ref{fig:images3} exemplifies that even for systems of comparable present-epoch mass, the technique is not generally applicable.  Fig. \ref{fig:images2} also shows an occasion where it appears satellites have remained after removing the identified substructure.  As we discuss in Appendix \ref{app:algorithm} and is exemplified in Fig. \ref{fig:images2}, the BaryMP radius successfully cuts inside the largest (star-dominated) satellite, as designed.

Conversely to Fig. \ref{fig:images1}, the cold gas in Fig. \ref{fig:images3} shows a fairly distinctive cut-off by eye, whereas the stars are dispersed more continuously throughout the subhalo.  Because BaryMP considers cold gas and stars equivalently, the technique is motivated to fairly compute an aperture radius in both instances, despite their differences.

\section{Results}
\label{sec:results}
We present results in a relative context, where measurements from each technique are normalised to the respective integrated properties of their full parent (sub)halo (\frenchspacing{i.e. where} no aperture has been used, while substructure and hot gas has been stripped).  We normalise to this technique not only because it has been popular in the literature to assume the properties of galaxies and their parent (sub)haloes are equivalent (\frenchspacing{cf. Table \ref{tab:techs}), but also as it shows what each aperture technique says about the baryons surrounding the galaxy. For relatively isolated systems (\frenchspacing{e.g. the M12 galaxies}), the baryons that occupy the outer regions of (sub)haloes represent streams of material and star clusters stripped from former satellites. For systems in dense environments, these particles are also representative of the observed intracluster light \citep[for details on intracluster light, see, e.g.,][]{gonzalez05}.

\subsection{Relative measurements for Milky-Way-mass systems}
\label{ssec:relm12}

We first address Milky-Way-mass systems by focusing on measurements from the M12 simulations with a supporting subsample of the \emph{MassiveBlack-II} galaxies.  This latter subsample includes subhaloes with stellar and gas masses each in the same range as the M12 simulations: $M_{\mathrm{stars}} \in [3,25]\times 10^{10}$ M$_{\odot}$, $M_{\mathrm{gas}} \in [0.85,8.5]\times 10^{10}$ M$_{\odot}$ (hot + cold).  In total, this provided 282 supporting subhaloes.

Figs. \ref{fig:radrel}--\ref{fig:sfrrel} present the primary results of this subsection.  Each of these figures contains three panels.  The (a) panels show the average of the relative measurements from the full suite of M12 simulations for each technique at each snapshot.  The (b) panels show the probability distribution functions for the M12 simulations for each technique for the snapshot nearest $z=0.0625$.  These distributions are generated using a Gaussian kernel density estimator, with a bandwidth obtained from Scott's rule \citep{scott}.  The (c) panels provide equivalently generated distributions from the \emph{MassiveBlack-II} Milky-Way-mass subsample.  Although some distributions may appear to continue outside the plotted $x$-axis range, it physically makes no sense for radii to be lower than 0, nor does it for any of the other properties to be outside $[0,1]$.\footnote{With the exception of gas mass for direct aperture techniques, as they count hot gas particles.  As such, those distributions were not renormalised. In practice, this has negligible impact on our results.}  As such, each distribution is normalised to have an area of 1 within the physically meaningful boundaries, and is effectively cut at these boundaries. Direct aperture techniques are represented by dashed curves while those that included a subhalo finder (and excluded hot gas) are given by solid curves.  Colours are associated with aperture techniques as defined in the legends.  Each of these figures is analysed in turn in the following subsidiary subsections.

We summarise results from this subsection and Section \ref{ssec:relmbii} in Table \ref{tab:dists}.  There, we provide the means and standard deviations for each of the datasets used to generate the distributions presented in Figs. \ref{fig:radrel}--\ref{fig:mbiirel}.

\subsubsection{Aperture radii}
\label{ssec:radrel}

\begin{figure*}[t]
\includegraphics[width=0.49\textwidth]{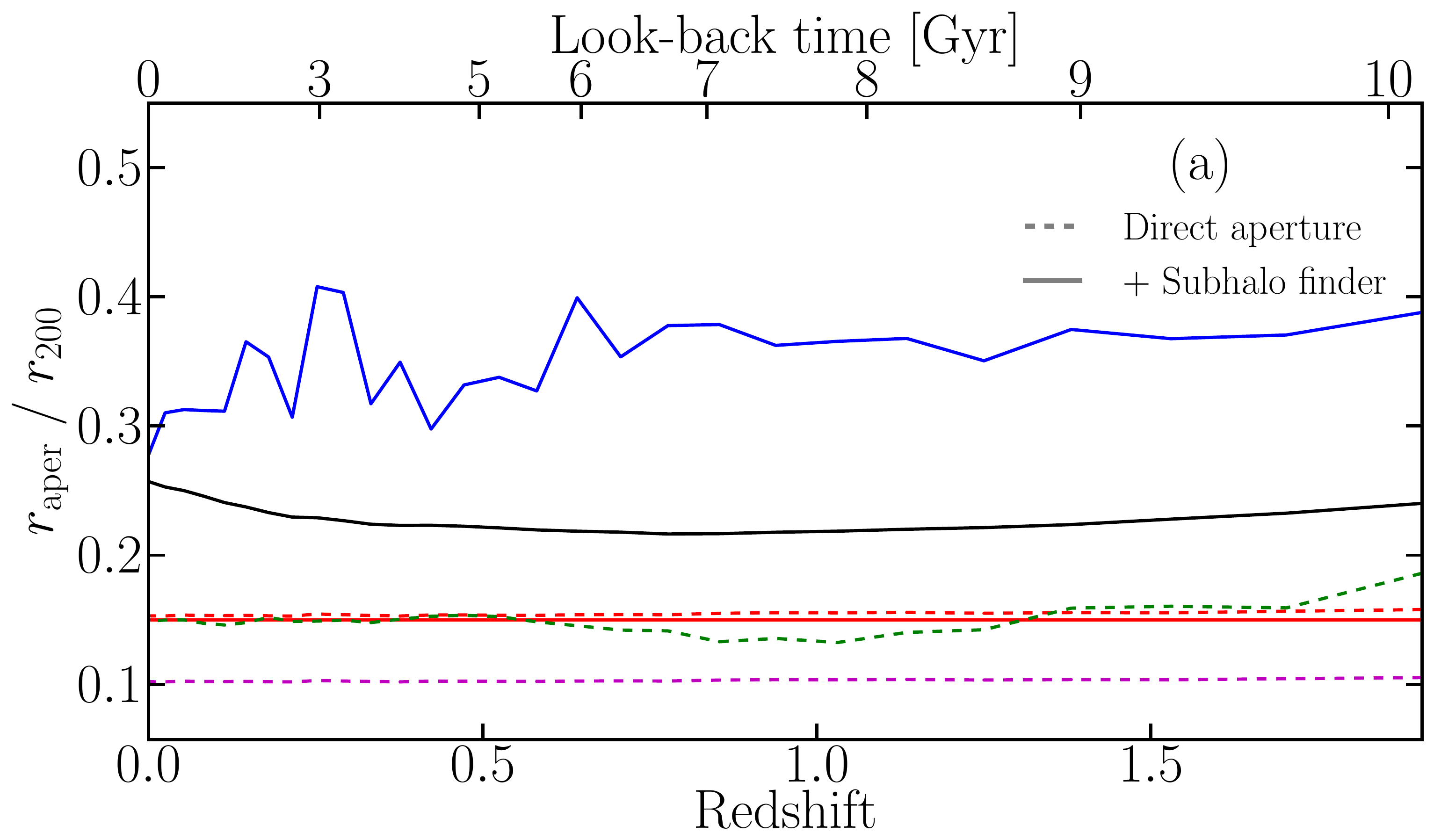}
\includegraphics[width=0.49\textwidth]{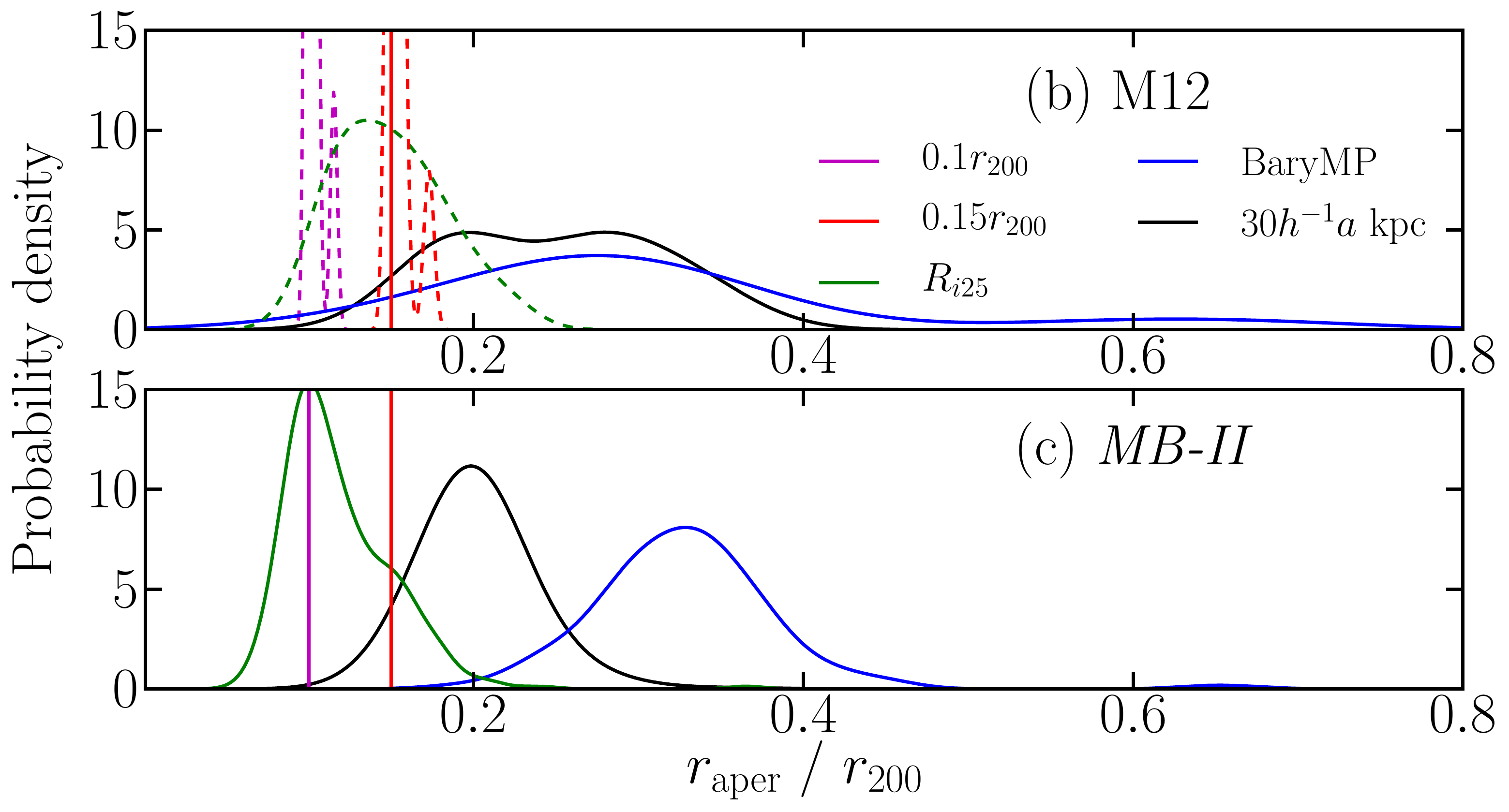}
\caption{Aperture radii for each technique normalised to the systems' virial radii.  Panel (a): Average of the M12 simulation sample at each redshift.  Panel (b): Probability distribution function generated from a Gaussian kernel density estimator for the normalised radii of the M12 galaxies at the snapshot nearest to $z=0.0625$.  Panel (c): Equivalent distribution for \emph{MassiveBlack-II} subhaloes at $z=0.0625$ for a subsample of 282 galaxies with comparable gas and stellar masses to the M12 galaxies.  Dashed lines represent the direct aperture techniques, while solid lines indicate the additional use of a subhalo finder and omission of hot gas.  Colours correspond to aperture techniques as follows: magenta = $0.1r_{200}$, red = $0.15r_{200}$, green = $R_{i25}$, black = $30h^{-1}a$ kpc, blue = BaryMP.  See Section \ref{ssec:radrel}.}
\label{fig:radrel}
\end{figure*}

To put the measurements of the integrated properties in context, the aperture radii from each technique are first presented in Fig. \ref{fig:radrel}. All measurements are normalised to $r_{200}$, calculated using only the particles identified by the relevant subhalo finder.

From Fig. \ref{fig:radrel}a, it is evident that, for the M12 simulations, the average $R_{i25}$ radius (green line) is practically equivalent to $0.15r_{200}$ (red lines), regardless of redshift.  Fig. \ref{fig:radrel}b shows that the data producing the $R_{i25} / r_{200}$ distribution come with $\sigma = 0.03$ at low redshift, though. The $0.15r_{200}$ technique hence provides a \emph{rough} approximation for the optical limit of these simulated galaxies.  The same can not quite be said for the \emph{MassiveBlack-II} systems however; while they carry the same variance for the $R_{i25} / r_{200}$ distribution, the mean is smaller (\frenchspacing{cf. Fig. \ref{fig:radrel}c}, Table \ref{tab:dists}).

From the blue distributions in Figs. \ref{fig:radrel}b and \ref{fig:radrel}c, we see again that BaryMP radii can cover a wide range of fractions of (sub)haloes' virial radii. Over all examined redshifts for the M12 galaxies, the mean of the $r_{\mathrm{BaryMP}} / r_{200}$ distribution does vary, but remains approximately between 0.3 and 0.4 (dropping below 0.3 at $z=0$, consistent with the discussion in Section \ref{ssec:barymp}; see Fig. \ref{fig:radrel}a).  

Comparison of Figs. \ref{fig:radrel}b and \ref{fig:radrel}c shows that the average M12-analogue in \emph{MassiveBlack-II} returns radii at lower fractions of $r_{200}$ for the fixed (comoving) aperture technique, while returning higher fractions for BaryMP.  The former point suggests the M12 galaxies occupy smaller haloes on average, while the latter indicates that the \emph{MassiveBlack-II} galaxies are less concentrated than the M12 ones.  The latter point is also supported by $R_{i25}$ values being larger for M12 systems.  We note that while the distribution for BaryMP in Fig. \ref{fig:radrel}b appears to reach radii of zero, there are no instances where this actually happens; this is merely a result of the extrapolative nature of kernel density estimation.

\subsubsection{Stellar mass}
\label{ssec:smrel}

\begin{figure*}[t!]
\includegraphics[width=0.49\textwidth]{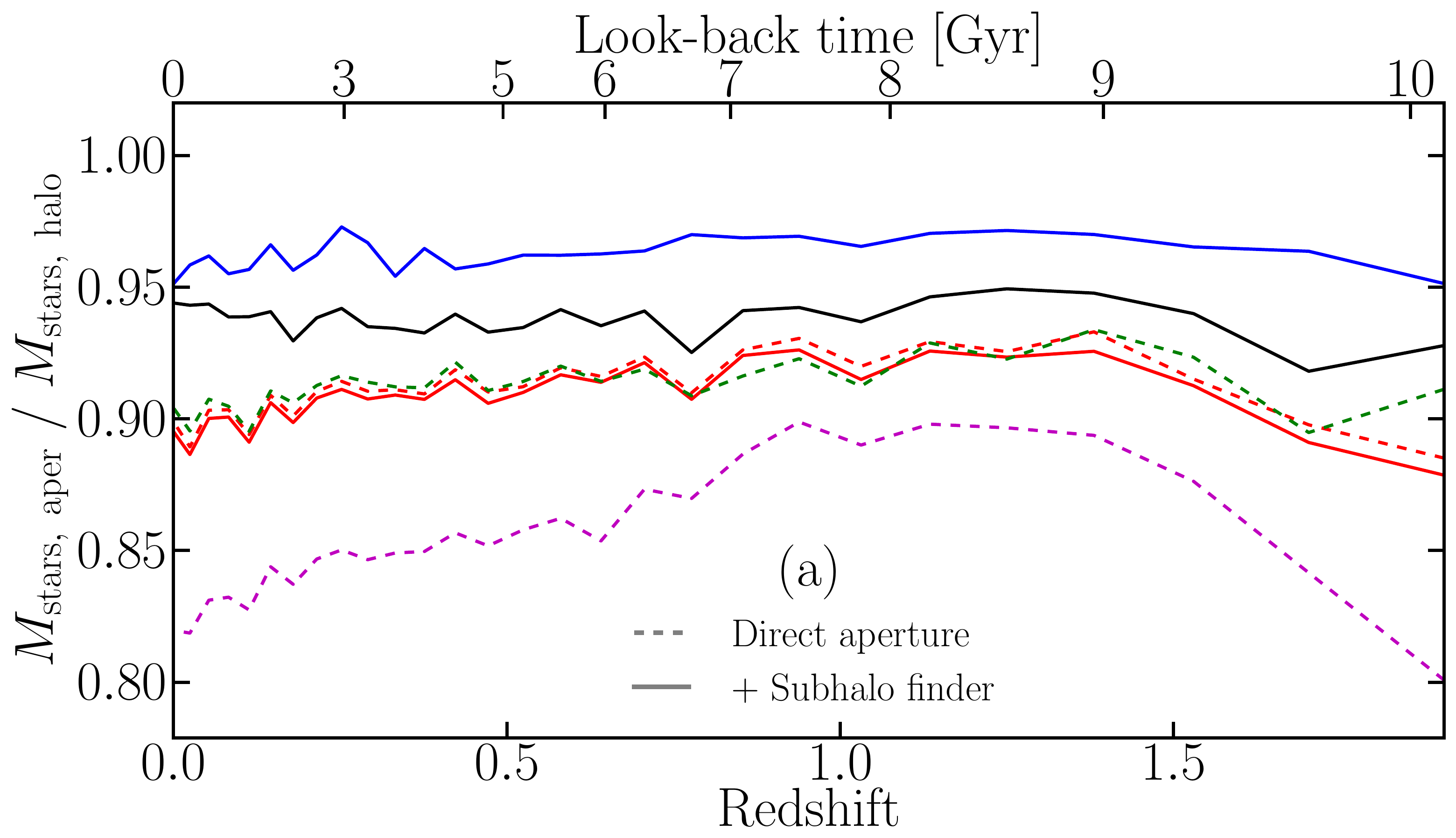}
\includegraphics[width=0.49\textwidth]{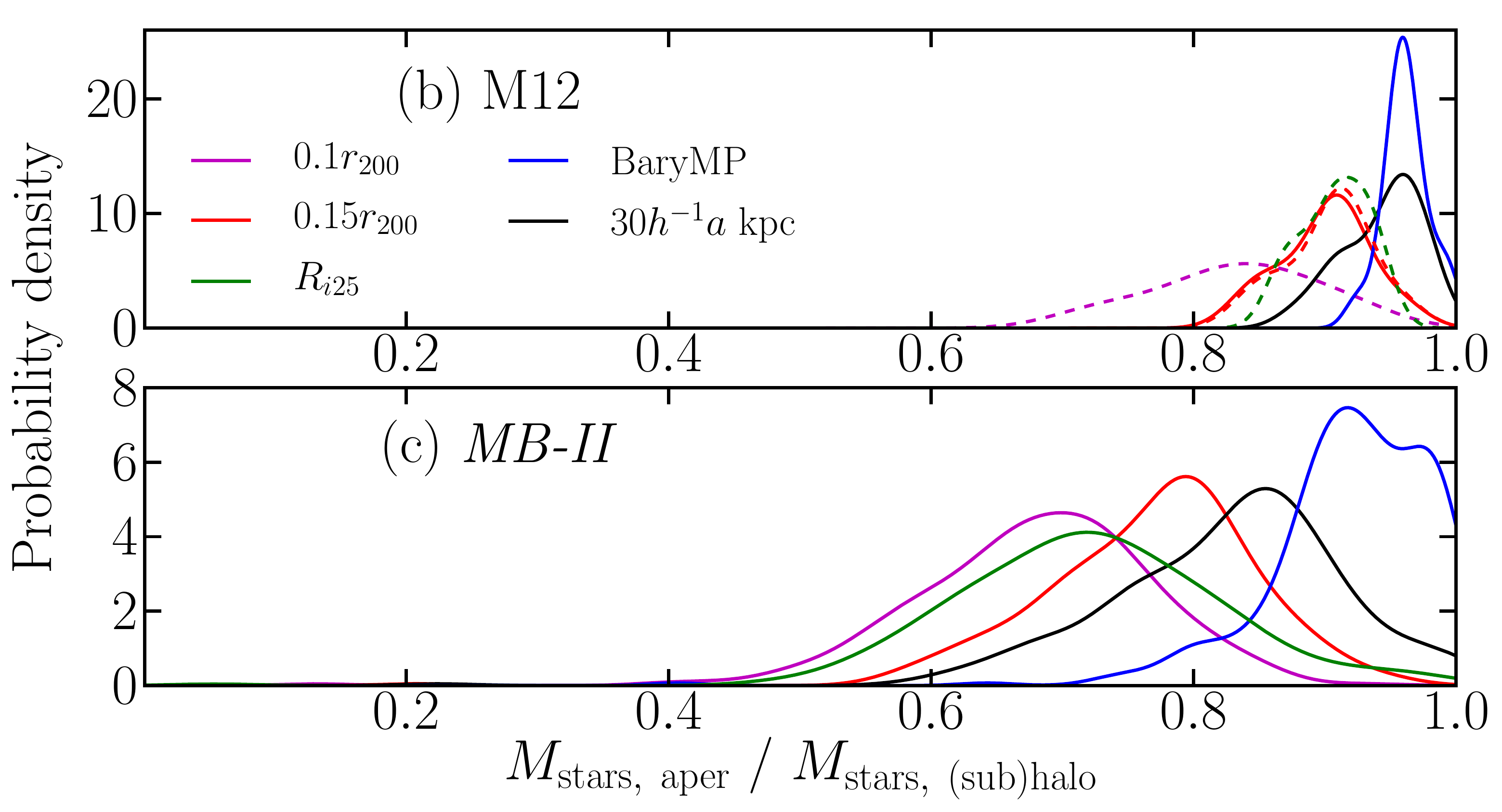}
\caption{Integrated stellar mass for each aperture technique normalised to the value for the full substructure-stripped (sub)halo.  Panel (a): Average values for the M12 systems as a function of redshift.  Panel (b): Probability distribution functions for the M12 galaxies for the snapshot closest to $z=0.0625$.  Panel (c): Equivalent distributions for the Milky-Way-mass \emph{MassiveBlack-II} systems. Differences between techniques are of order tens of per cent, and can be as large as a factor of 2.  Further details in Section \ref{ssec:smrel}.  Plotting conventions maintained from Fig. \ref{fig:radrel}.}
\label{fig:smrel}
\end{figure*}

Stellar mass measurements are presented in Fig. \ref{fig:smrel}. Over all redshifts, the $0.15r_{200}$ and $R_{i25}$ apertures, on average, predict stellar masses $\sim$10 per cent less than that of the full halo for the M12 simulations.  The $0.1r_{200}$ technique naturally shows a larger average difference, nearing 20 per cent at $z = 0$ and 2.  The \emph{MassiveBlack-II} systems return lower average stellar masses for $0.1r_{200}$, $0.15r_{200}$, and $30h^{-1}a$ kpc than those of M12, consistent with the M12 galaxies being more concentrated (and in smaller haloes).  The shapes of the distributions for these techniques are almost identical for the different simulations, though (\frenchspacing{cf. Figs. \ref{fig:smrel}b and \ref{fig:smrel}c}).

Comparing the two most popular techniques in the literature ($0.1r_{200}$ and no aperture; \frenchspacing{cf. Table} \ref{tab:techs}), an average \emph{MassiveBlack-II} galaxy shows a difference in stellar mass of 32 per cent.  At the lower extreme of the relevant distribution (magenta, Fig. \ref{fig:smrel}c), the potential for a factor-of-two difference in stellar mass is seen.

For the M12 simulations, the measured stellar mass from BaryMP is never more than a few per cent less than the full halo.  The larger sample of \emph{MassiveBlack-II} galaxies indicates there is still the potential to find up to a 30-per-cent difference with the application of BaryMP versus the full subhalo, however (lower extreme, blue distribution, Fig. \ref{fig:smrel}c).

\subsubsection{Gas mass}
\label{ssec:gmrel}

\begin{figure*}[t!]
\includegraphics[width=0.49\textwidth]{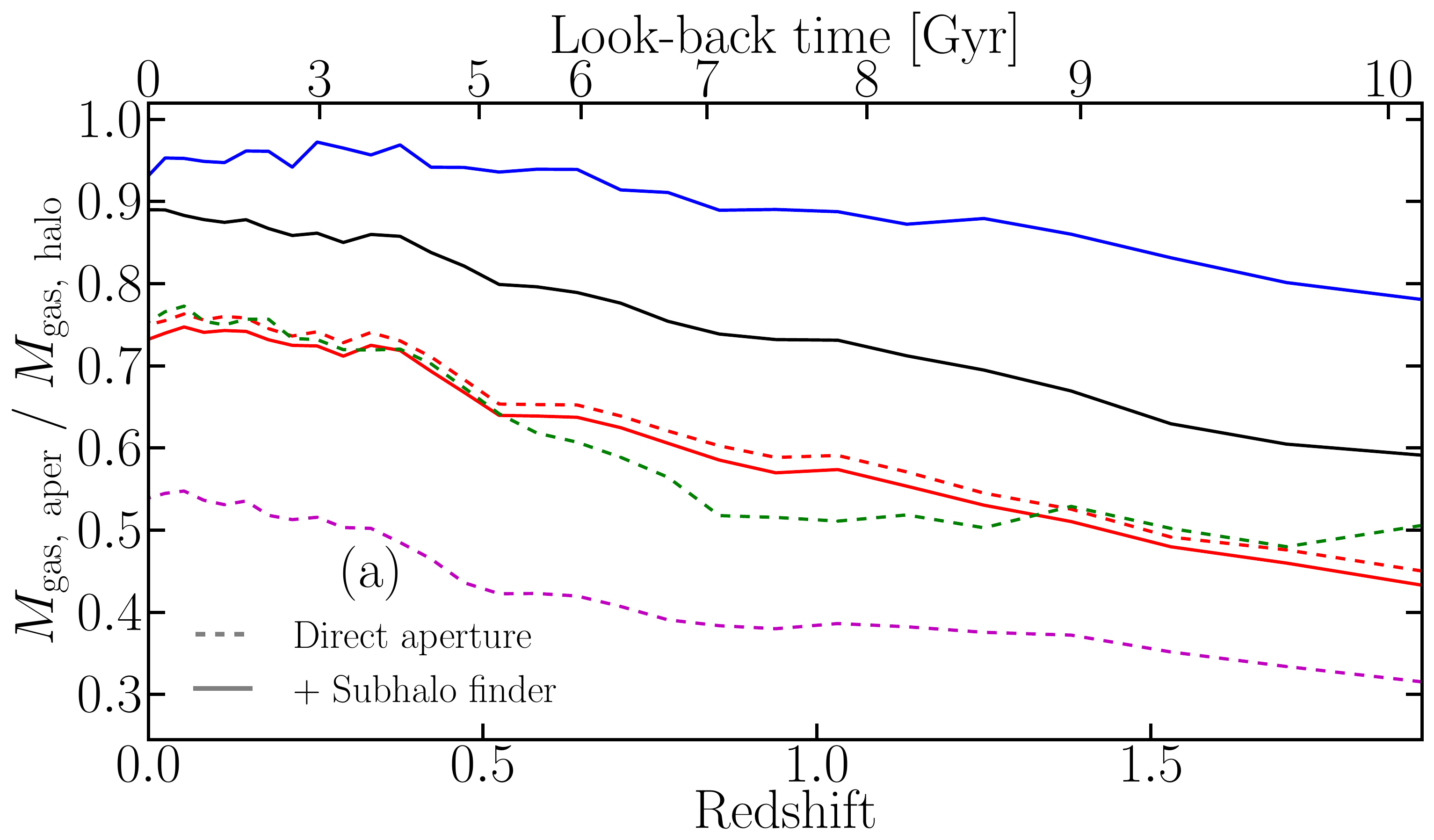}
\includegraphics[width=0.49\textwidth]{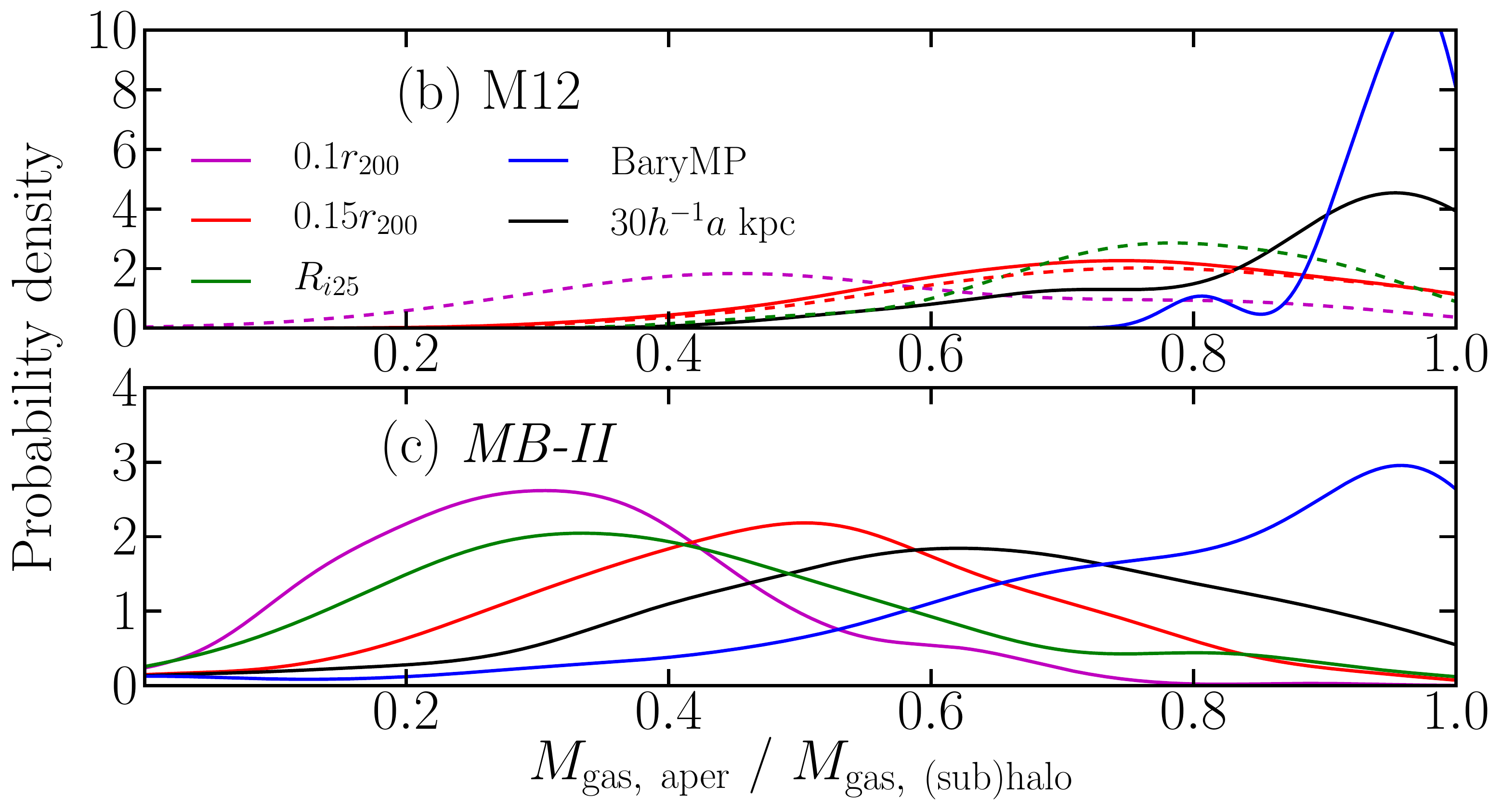}
\caption{As for Fig. \ref{fig:smrel} but displaying the integrated gas mass instead. For techniques that used a subhalo finder, only cold gas particles contributed, while direct aperture techniques included hot gas.  Large differences are evident from the various apertures, with the potential to exclude cold gas entirely.  More analysis in Section \ref{ssec:gmrel}.}
\label{fig:gmrel}
\end{figure*}

The radial density gradient of gas in galaxies is typically noticeably shallower than it is for stars, especially towards their centres.\footnote{This is a natural consequence of stars forming from gas and only doing so in the densest regions.  Visually, this can be readily seen from Fig. \ref{fig:images1} (but also from Figs. \ref{fig:images2} and \ref{fig:images3}), noting the difference in intensity scale given in the caption.  This has also been empirically shown to be the case with H\thinspace\textsc{i} + H$_2$ observations \citep[e.g., see appendix F of][]{leroy08}.}  As such, there is a greater variation in measured gas mass from each technique, as shown in Fig. \ref{fig:gmrel}.  Comparing the application of $0.1r_{200}$ and no aperture (the two most popular and yet most contrasting techniques), for the average M12 simulation, there is factor of $\sim$3 difference in the measured gas mass at high redshift, coming closer to 2 at low redshift.  At $z \sim 0$, the greatest individual difference for the M12 galaxies is a factor of 5, while 23 per cent of \emph{MassiveBlack-II} galaxies show differences at least this large (see magenta distributions in Figs. \ref{fig:gmrel}b and \ref{fig:gmrel}c, respectively).

As for other properties, the BaryMP gas masses do not deviate excessively from the full M12 halo values (blue distribution, Fig. \ref{fig:gmrel}b).  The \emph{MassiveBlack-II} systems show a much wider range of BaryMP gas masses, however, with cases beyond a factor-of-two difference to the full subhaloes.  Fig. \ref{fig:gmrel}a also shows the aperture to play a larger role at higher redshift.

\subsubsection{Star formation rate}
\label{ssec:sfrrel}

\begin{figure*}[t!]
\includegraphics[width=0.49\textwidth]{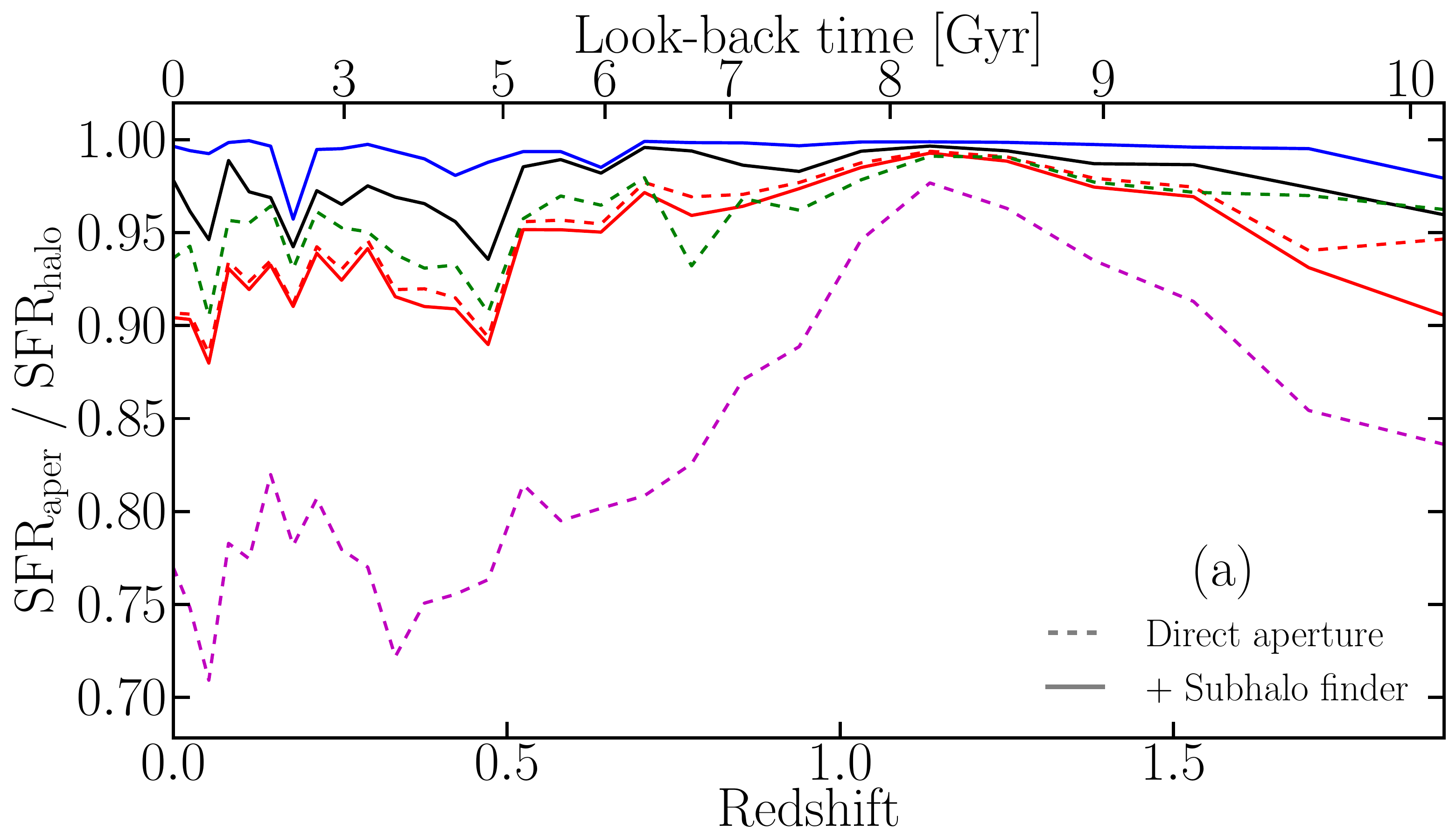}
\includegraphics[width=0.49\textwidth]{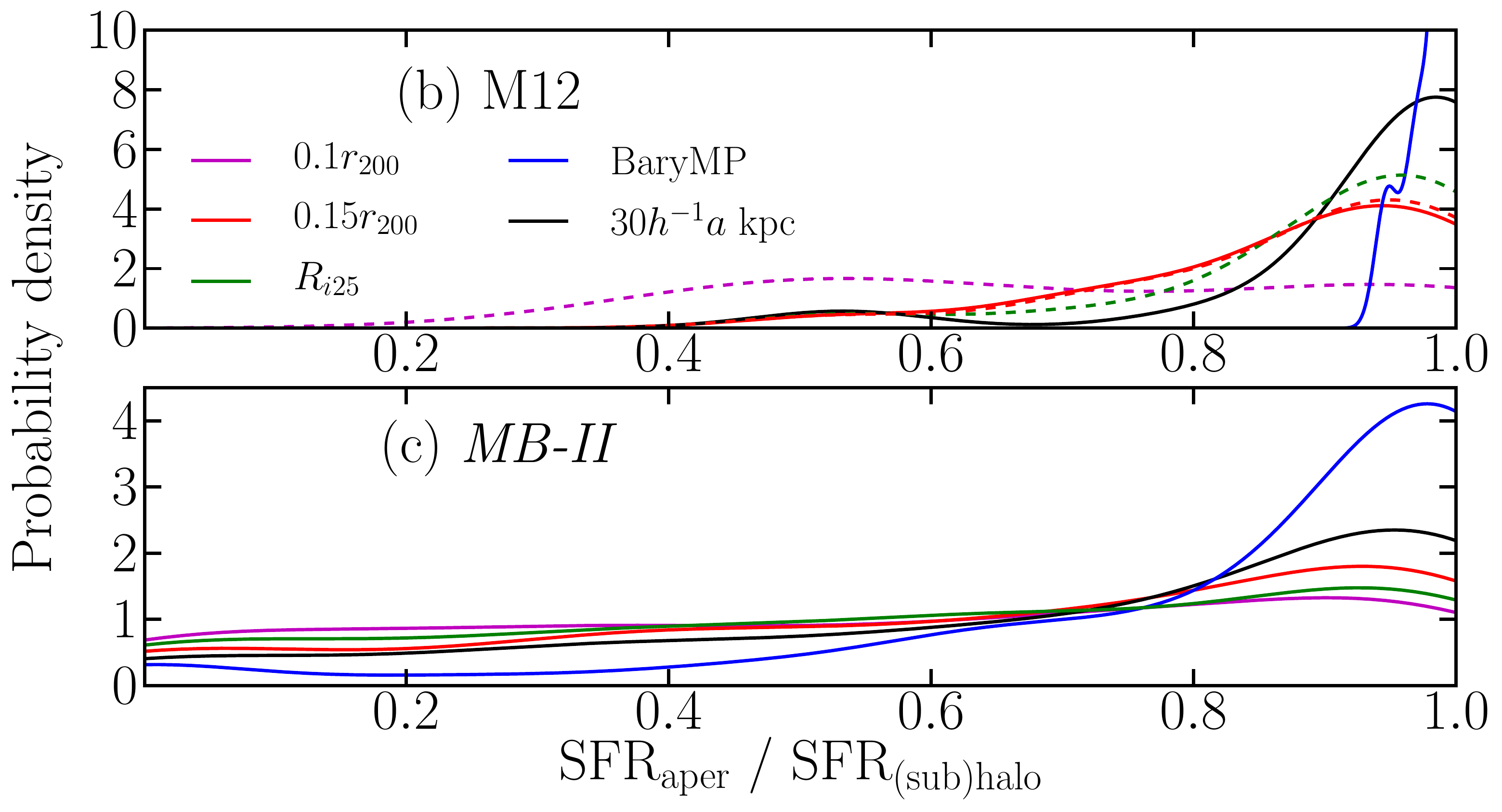}
\caption{As for Fig. \ref{fig:smrel} but displaying the integrated star formation rate instead.  As with gas mass, in the more extreme cases, the apertures can exclude the star-forming region entirely for the \emph{MassiveBlack-II} systems (see Section \ref{ssec:sfrrel}).}
\label{fig:sfrrel}
\end{figure*}

The integrated star formation rate measurements for the M12 galaxies, displayed in Fig. \ref{fig:sfrrel}, were calculated by summing the mass of all star particles identified by a given technique that were also identified as being gas particles in the galaxy in the previous snapshot, and dividing this value by the time-step. In truth, this means each value is the average star formation rate over the 375 Myr prior. At high redshift, most of the star formation occurs in a small, central region, meaning most techniques return similar values. With the formation of discs, the star-forming region extends, with $>$20 per cent of star formation within the halo occurring beyond $0.1r_{200}$ for the average M12 system.  For individual systems, values are shown to vary by up to and above 60 per cent for different techniques for the M12 galaxies (magenta distribution, Fig. \ref{fig:sfrrel}b).

Following \citet{springelhernquist}, each gas particle in \emph{MassiveBlack-II} has a tracked star formation rate.  Integrated values for each galaxy were calculated by summing these.  In truth, this provides the forecasted star formation rate over the next simulation time-step, rather than the rate at which stars actually formed in the end.  Regardless of the differing definition to the M12 galaxies, each technique (with perhaps the exception of BaryMP) shows similar distributions for the two simulation sets.  For the majority of these \emph{MassiveBlack-II} galaxies, star formation is confined to a more central region, leading to star formation rate measurements showing less susceptibility to aperture technique than gas mass.  

In some cases for \emph{MassiveBlack-II}, however, the choice of aperture can exclude the star-forming region entirely.  We note that if a system had no star formation within the entire subhalo, that system did not contribute to any of the distributions presented in Fig. \ref{fig:sfrrel}c.

\subsubsection{Gas accretion and ejection}
\label{ssec:gtrrel}

\begin{figure*}[t!]
\centering
\includegraphics[width=0.49\textwidth]{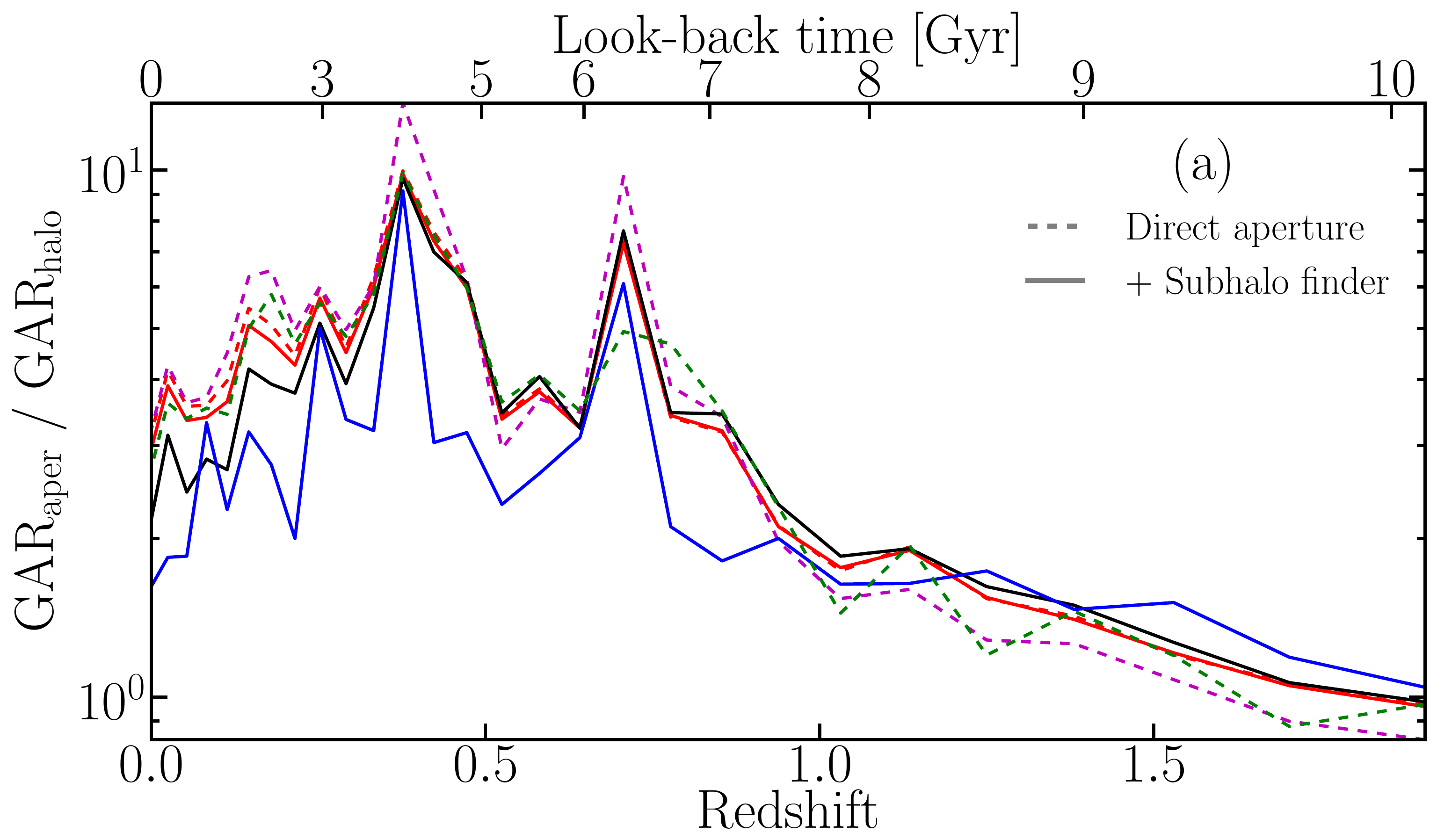}
\includegraphics[width=0.49\textwidth]{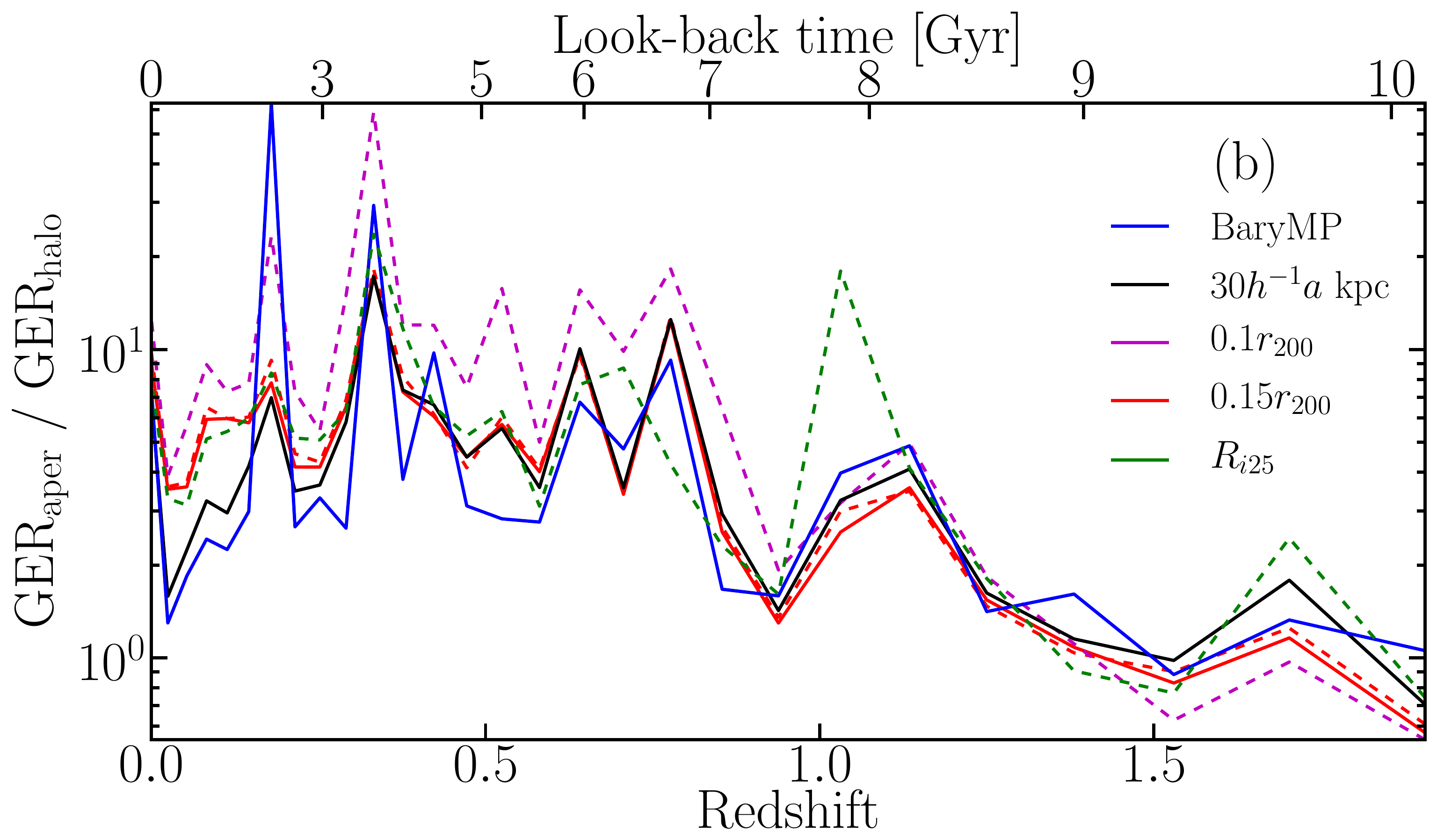}
\caption{Average gas accretion rate (a) and gas ejection rate (b) for the M12 galaxies. Curves as for the (a) panels of Figs. \ref{fig:smrel}--\ref{fig:sfrrel}.  These properties show the greatest susceptibility to aperture technique choice (elaborated in Section \ref{ssec:gtrrel}), as indicated by the logarithmic $y$-axes; note that these make the dashed and solid curves for $0.15r_{200}$ scarcely separable though.}
\label{fig:gtrrel}
\end{figure*}

Two properties that show an even greater dependence on measurement technique are the gas accretion and ejection rates. We present relative measurements for the M12 galaxies in Fig. \ref{fig:gtrrel}, where the respective rates were determined by tracking which gas particles were considered part of the galaxy in temporally adjacent snapshots. Once again, only cold gas particles were counted for the techniques that used \textsc{ahf}. No separation between steady accretion and mergers was made. The order of which techniques give the largest differences remains similar to the other properties at high redshift, but is reversed at low redshift for these plots. This is consistent with the notion that, at early epochs, haloes accumulate cold gas rapidly from their surroundings, where much of this gas penetrates to the central galaxy in cold streams.  Shortly after, as the haloes' masses increase, the majority of gas is shock-heated and sits in the halo \citep{reesos}. At lower redshift, less gas shifts through the boundary of the virial sphere, while the gas inside it continues to cool onto the galaxy, and feedback mechanisms cause outflows which repopulate the halo gas content. As a rule of thumb, at low redshift, the closer to the centre of the galaxy one defines a surface, the faster the gas will be measured to pass through that surface in both directions. Differences of up to and above an order of magnitude in gas transfer rates emphasize the importance of how the end of galaxies is defined.  The technique-to-technique difference varies strongly from galaxy to galaxy (see Table \ref{tab:dists}).  A study of the gas accretion and ejection rates in \emph{MassiveBlack-II} systems fell beyond the scope of the project.

\subsection{Relative measurements for a broad galaxy population}
\label{ssec:relmbii}

\begin{figure*}[t!]
\centering
\includegraphics[width=0.49\textwidth]{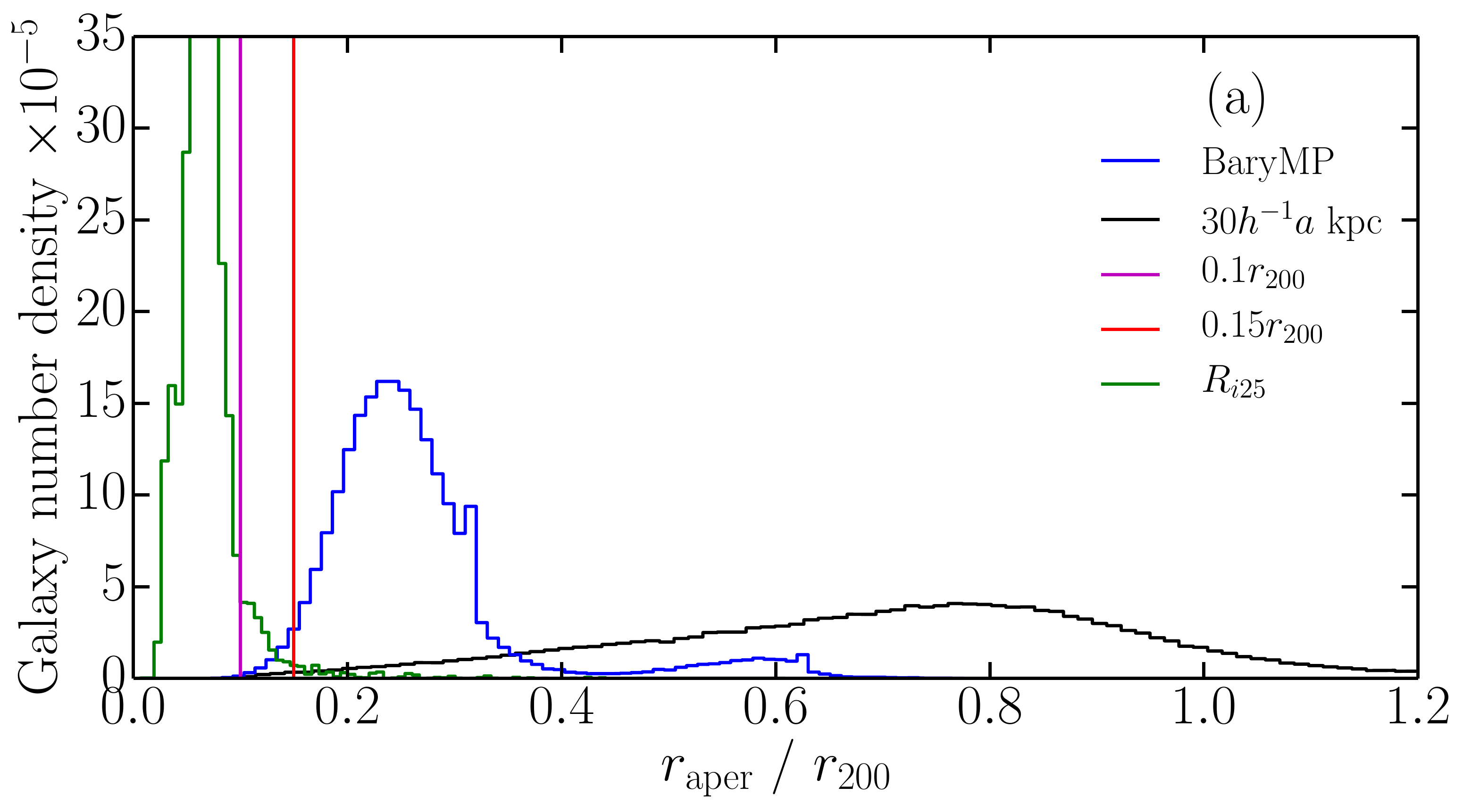}
\includegraphics[width=0.49\textwidth]{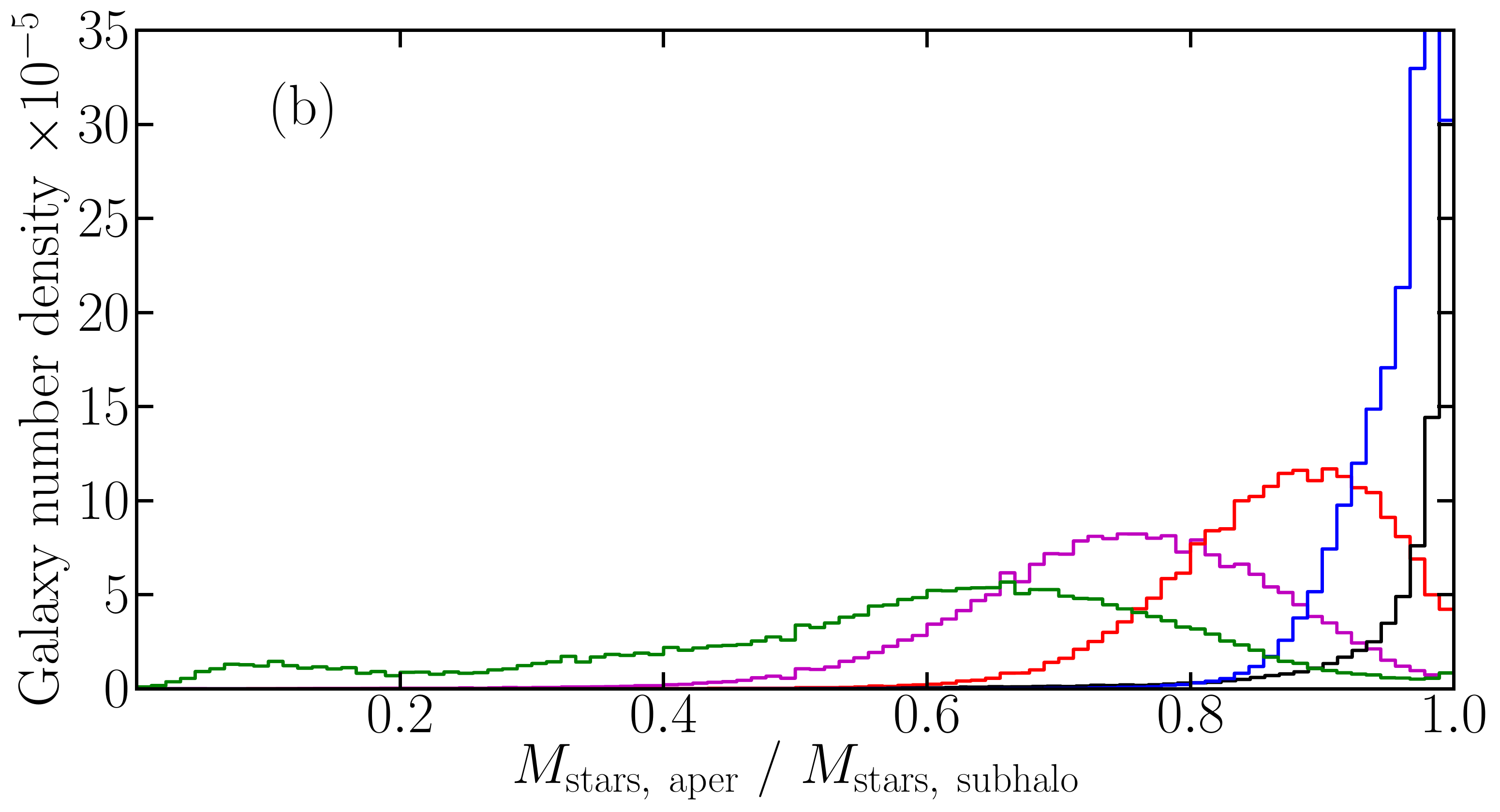}
\includegraphics[width=0.49\textwidth]{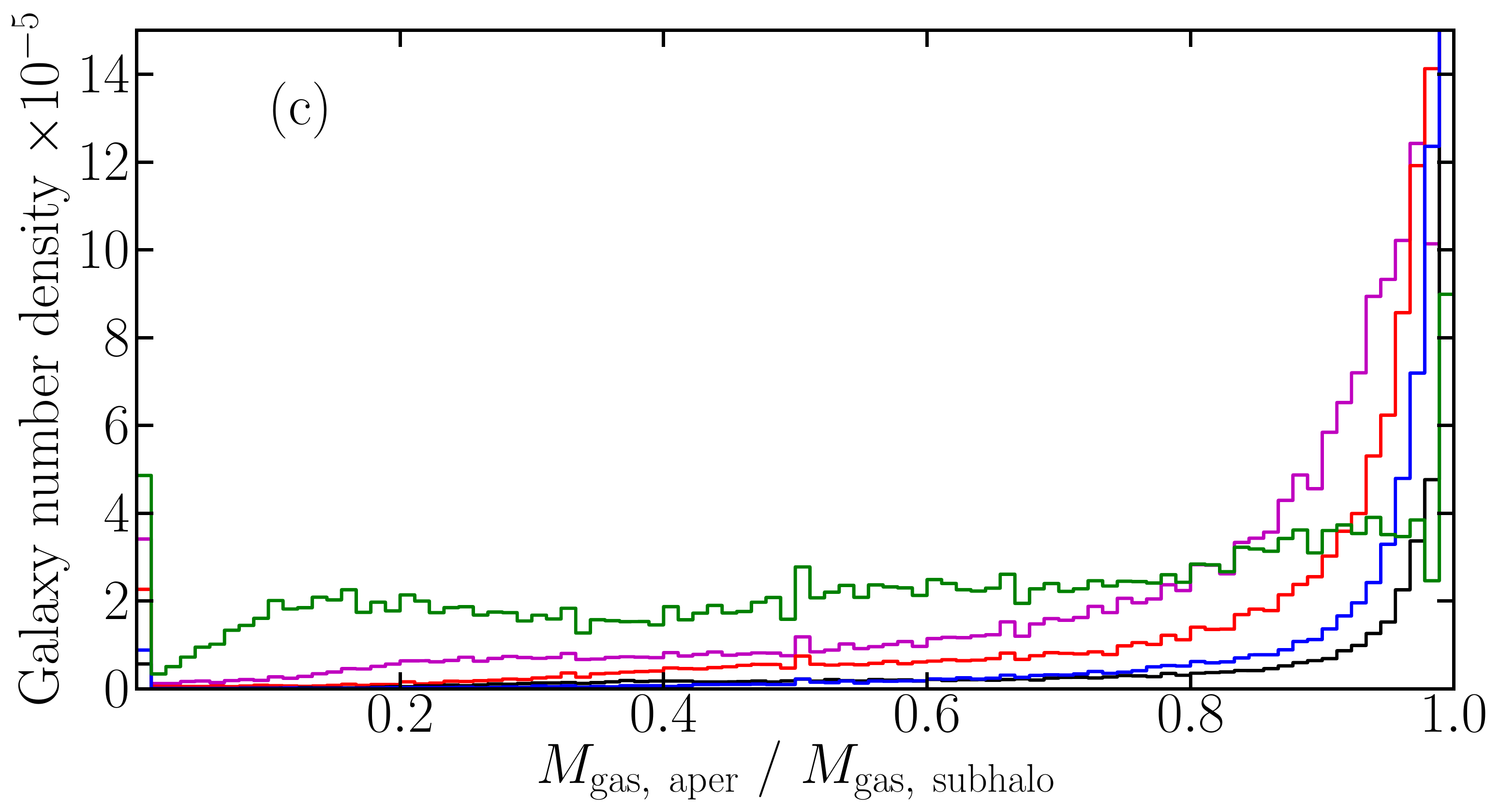}
\includegraphics[width=0.49\textwidth]{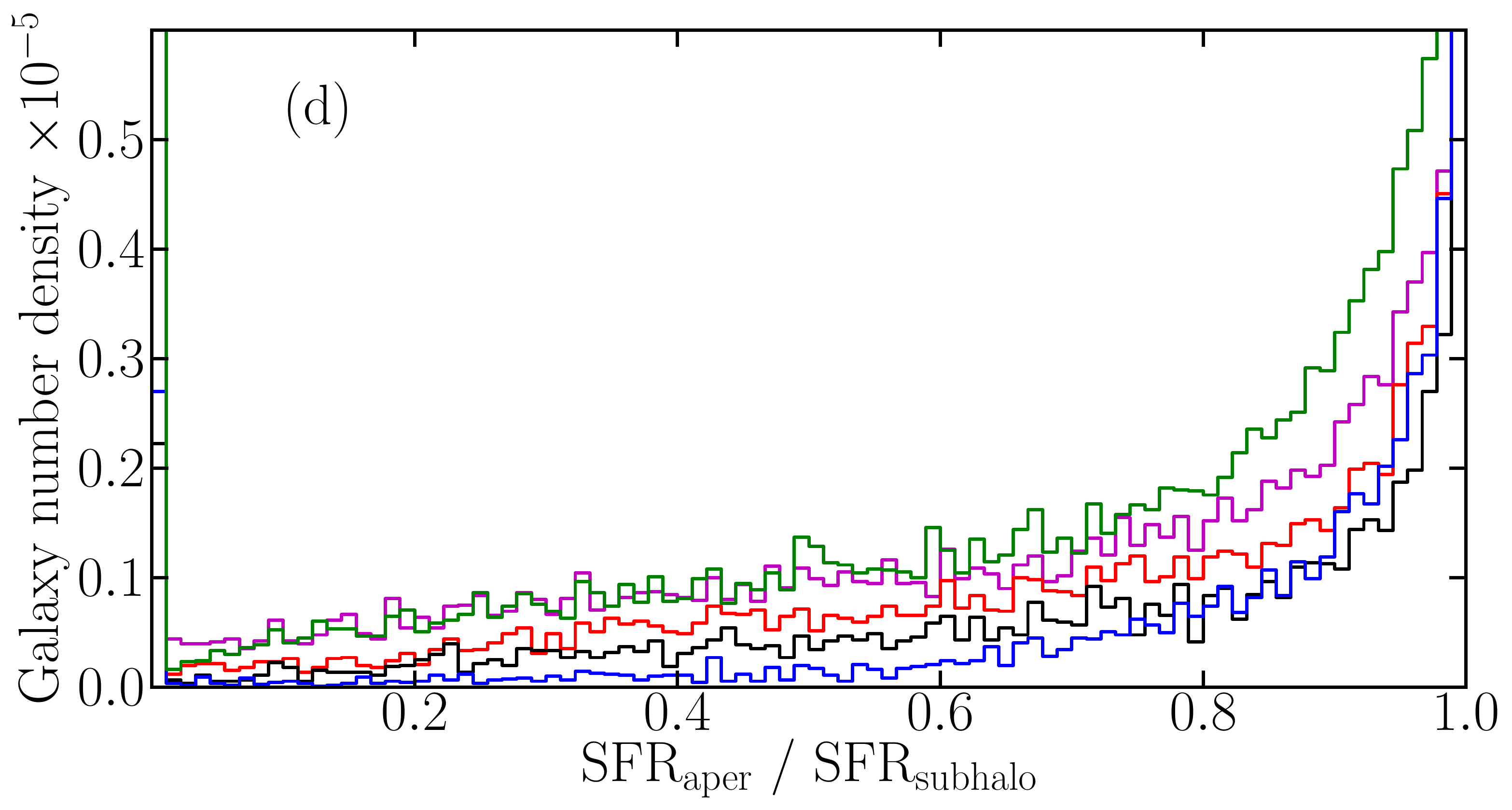}
\caption{Integrated properties of a general galaxy population (covering 4 orders of magnitude in stellar and gas mass) relative to the properties of their parent subhaloes: (a) aperture radius, (b) stellar mass, (c) gas mass, (d) star formation rate. Each panel is similar to the (b) panels of Figs. \ref{fig:radrel}--\ref{fig:sfrrel} but covers the full \emph{MassiveBlack-II} sample ($2.2 \times 10^5$ galaxies) and shows the actual binned values.  With the exception of stellar mass, overall, this broader galaxy population shows lower aperture--dependence on integrated-property measurements (analysis in Section \ref{ssec:relmbii}).  Note that the $y$-axis in panel (d) is zoomed in significantly further than the others.}
\label{fig:mbiirel}
\end{figure*}

We present results for the full \emph{MassiveBlack-II} sample with four plots in Fig. \ref{fig:mbiirel}.  All comments in this subsection are made with respect to this figure.

The range of radii relative to the virial radius for the BaryMP technique for this large sample of simulated galaxies is consistent with the Milky-Way-mass sample, indicating that the technique is unbiased toward subhalo baryonic mass (cf. blue distributions in Figs. \ref{fig:radrel}c and \ref{fig:mbiirel}a).  The hugely variant size of subhaloes in this sample leads to the fixed aperture returning a very wide distribution when normalised to the virial radius (black histogram, Fig. \ref{fig:mbiirel}a), with the aperture larger than $r_{200}$ itself on some occasions, highlighting the inappropriateness of broadly applying such a technique.  While $R_{i25}$ radii are comparable to $0.1r_{200}$ for the Milky-Way-mass systems, the optical limit for the general \emph{MassiveBlack-II} galaxy is much smaller (cf. green distributions in Figs. \ref{fig:radrel}c and \ref{fig:mbiirel}a).

Stellar mass measurements, presented in Fig. \ref{fig:mbiirel}b, show a relatively strong dependence on technique, with $0.1r_{200}$ averaging 25 per cent less mass than the full subhalo, with the tail of the magenta distribution maintaining a potential for differences of 50 per cent.  $R_{i25}$ stellar masses show even more extreme cases, where almost all the star particles in the subhalo can be excluded by the aperture.  The distributions from $0.1r_{200}$ and $0.15r_{200}$ exhibit similarities to the Milky-Way-mass sample, but with higher averages (cf. Figs. \ref{fig:smrel}c and \ref{fig:mbiirel}b).  Given that the small, low-mass systems vastly outnumber the high-mass ones, the largeness of an aperture of radius $30h^{-1}a$ kpc leads to stellar masses barely smaller than the full subhalo, on average.  Meanwhile, BaryMP typically returns stellar masses 4 per cent lower than the full subhalo, but can be 20 per cent lower in the extreme cases.

Conversely, with the exception of $R_{i25}$, gas mass shows relatively little variation with technique (Fig. \ref{fig:mbiirel}c) for this broader sample.  This suggests that the cold gas in the average galaxy is more tightly concentrated than in Milky-Way-mass systems.  Still, 30 per cent of galaxies for $0.1r_{200}$ and 14.5 per cent for $0.15r_{200}$ give gas masses $\geq$20 per cent less than when no aperture is used, while half the subhaloes have at least 37 per cent of their cold gas excluded by the $R_{i25}$ aperture.  There is also a noticeable population of galaxies with gas masses of zero for all the techniques.  In these cases, the gas in the subhalo is offset from the stellar centre.  Processes such as ram-pressure stripping can be held responsible for this.

The star formation rate distributions are effectively scaled-down versions of cold gas mass (Fig. \ref{fig:mbiirel}d); only the \emph{very} dense gas can form stars.  This property is scarcely subject to technique for galaxies in general, with, for example, only 4 per cent of the sample returning values $\geq$5 per cent less than the \textsc{subfind} results for the $0.1r_{200}$ aperture.  There remains a small population of systems with measured star formation rates of zero from several techniques, corresponding to the subhaloes where the (star-forming) gas is off-centre.  We note again that systems where the entire subhalo had no star formation were omitted from this analysis (\frenchspacing{i.e. there is star} formation occurring within the subhaloes but outside the apertures for those examples).

\begin{table*}[t]
\centering
\begin{tabular}{c l c c c c c c c c c c}
\hline
Property & Sample & \multicolumn{2}{c}{$0.1r_{200}$} & \multicolumn{2}{c}{$0.15r_{200}$} & \multicolumn{2}{c}{$R_{i25}$} & \multicolumn{2}{c}{$30h^{-1}a$ kpc} & \multicolumn{2}{c}{BaryMP}\\
&& $\mu$ & $\sigma$ & $\mu$ & $\sigma$ & $\mu$ & $\sigma$ & $\mu$ & $\sigma$ & $\mu$ & $\sigma$\\
\hline

\multirow{3}{*}{$\dfrac{r_{\mathrm{aper}}}{r_{200}}$} & M12 & .102 & .004 & 0.15 & ------ & 0.15 & 0.03 & 0.25 & 0.06 & 0.31 & 0.13\\
& \emph{MB-II} MW & 0.10 & ------ & 0.15 & ------ & 0.12 & 0.03 & 0.21 & 0.09 & 0.33 & 0.06\\
& \emph{MB-II} all & 0.10 & ------ & 0.15 & ------ & 0.07 & 0.03 & 0.74 & 0.30 & 0.27 & 0.10\\

\multirow{3}{*}{$\dfrac{M_{\mathrm{stars,\ aper}}}{M_{\mathrm{stars,\ (sub)halo}}}$} & M12 & 0.83 & 0.06 & 0.90 & 0.03 & 0.91 & 0.02 & 0.94 & 0.03 & 0.96 & 0.02\\
& \emph{MB-II} MW & 0.68 & 0.09 & 0.77 & 0.08 & 0.72 & 0.10 & 0.83 & 0.09 & 0.92 & 0.07\\
& \emph{MB-II} all & 0.75 & 0.12 & 0.86 & 0.08 & 0.58 & 0.21 & 0.99 & 0.04 & 0.96 & 0.04\\

\multirow{3}{*}{$\dfrac{M_{\mathrm{gas,\ aper}}}{M_{\mathrm{gas,\ (sub)halo}}}$} & M12 & 0.55 & 0.20 & 0.75 & 0.16 & 0.77 & 0.12 & 0.88 & 0.13 & 0.95 & 0.05\\
& \emph{MB-II} MW & 0.32 & 0.15 & 0.49 & 0.19 & 0.40 & 0.21 & 0.62 & 0.22 & 0.79 & 0.22\\
& \emph{MB-II} all & 0.81 & 0.24 & 0.91 & 0.18 & 0.59 & 0.30 & 0.97 & 0.11 & 0.97 & 0.10\\

\multirow{3}{*}{$\dfrac{\mathrm{SFR}_{\mathrm{aper}}}{\mathrm{SFR}_{\mathrm{(sub)halo}}}$} & M12 & 0.70 & 0.22 & 0.88 & 0.13 & 0.91 & 0.12 & 0.95 & 0.12 & 0.99 & 0.02\\
& \emph{MB-II} MW & 0.56 & 0.33 & 0.64 & 0.32 & 0.59 & 0.33 & 0.71 & 0.31 & 0.83 & 0.26\\
& \emph{MB-II} all & 0.95 & 0.17 & 0.97 & 0.13 & 0.95 & 0.19 & 0.99 & 0.08 & 0.99 & 0.06\\

GAR$_{\mathrm{aper}}$ / GAR$_{\mathrm{(sub)halo}}$ & M12 & 3.57 & 2.39 & 3.35 & 2.52 & 3.38 & 2.16 & 2.45 & 1.78 & 1.85 & 1.04\\
GER$_{\mathrm{aper}}$ / GER$_{\mathrm{(sub)halo}}$ & M12 & 5.67 & 7.96 & 3.59 & 4.17 & 3.13 & 3.18 & 2.24 & 2.35 & 1.84 & 2.03\\

\hline
\end{tabular}
\caption{Summary of the datum samples that produced the distributions presented in Figs. \ref{fig:radrel}--\ref{fig:mbiirel}, i.e.~for the $z=0.0625$ or nearest available snapshots.  Mean, $\mu$, and standard deviation, $\sigma$, values are provided for each technique applied to each simulated-galaxy sample for each property.  Each aperture has been applied with a subhalo finder, except in the cases of $0.1r_{200}$ and $R_{i25}$ for the M12 systems (as per Table \ref{tab:apertures}).  Non-applicable entries are filled with horizontal lines. These values can be used to quantify, with uncertainties, how different techniques compare at measuring galaxy properties.  We note that the $\mu$ and $\sigma$ values for the distributions presented in Figs. \ref{fig:radrel}--\ref{fig:sfrrel} vary slightly ($<$10 per cent) to the values given here, which use the actual data.}
\label{tab:dists}
\end{table*}

\subsection{Galaxy scaling relations}
\label{ssec:scale}

Observationally, galaxy properties are known to follow certain scaling relations.  Often such relations can provide tests for how well a simulation has reproduced the real Universe.  We find it informative to check whether a simulation's ability to match these relations depends on the technique choice for measuring the relevant galaxy properties.

\subsubsection{Stellar mass--star formation rate relation}
\label{ssec:sfr}

A redshift-dependent power-law relation has been observed between stellar mass and star formation rate \citep[e.g.][]{daddi07,elbaz07,noeske07}.  Studying blue galaxies at low redshift, with a linear fit in log-log space, \citet{elbaz07} showed
\begin{equation}
\label{eq:smsfr}
\frac{\mathrm{SFR}}{\mathrm{M}_{\odot}\ \mathrm{yr}^{-1}} = 8.7^{+7.4}_{-3.7} \left(\frac{M_{\mathrm{stars}}}{10^{11}\ \mathrm{M}_{\odot}}\right)^{0.77},\ 0.015 < z < 0.1\ .
\end{equation}
\noindent We compare stellar masses and star formation rates for each aperture technique listed in Table \ref{tab:apertures} (with subhalo finders applied where listed) with Fig. \ref{fig:smsfr}, where contours represent the \emph{MassiveBlack-II} systems and dots the M12 galaxies.  The two M12 snapshots in the quoted redshift range of Equation \ref{eq:smsfr} were considered simultaneously.  

\begin{figure}[t!]
\centering
\includegraphics[width=\textwidth]{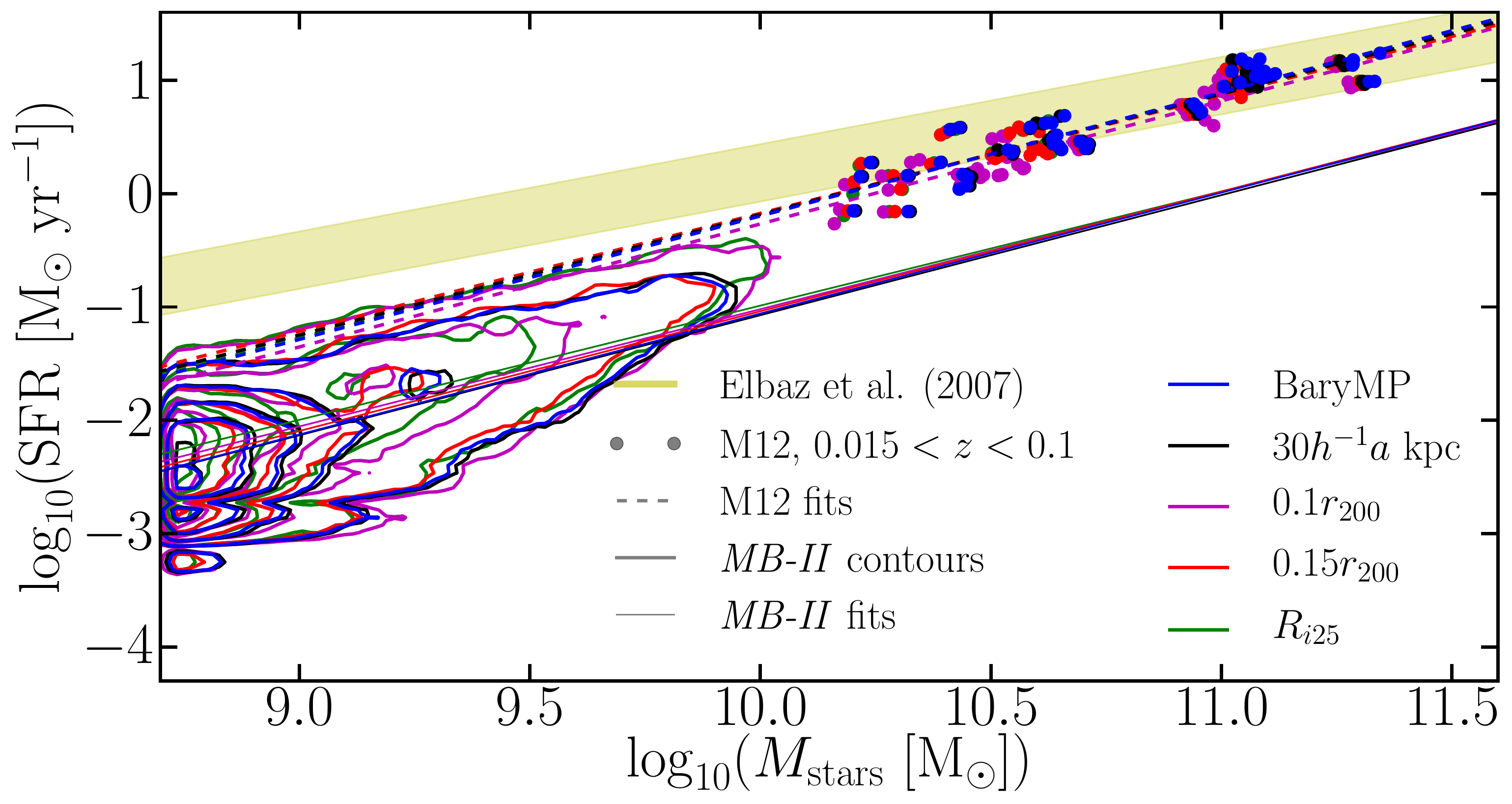}
\caption{Stellar mass--star formation rate relation for each aperture technique applied with a subhalo finder where possible (see Table \ref{tab:apertures}). The distributions of \emph{MassiveBlack-II} galaxies are shown with contours.  Individual dots represent the M12 galaxies for each of the two snapshots that fall in the redshift range of Equation \ref{eq:smsfr}, while the equation itself is shown by the yellow shaded region.  Aperture techniques follow the colour-coding convention of the previous figures.  Displayed straight-line fits to each dataset show a lack of dependence on aperture technique to reproduce this scaling relation.}
\label{fig:smsfr}
\end{figure}

For each technique, the M12 simulations show good agreement with the range suggested by Equation \ref{eq:smsfr}, with root-mean-squared (RMS) scatter values $<$0.27 dex from the centre of the yellow strip.  The best-fitting slopes from these data alone vary little from aperture to aperture, but do differ from the \citet{elbaz07} value, with values between 1.04 and 1.09.

For the \emph{MassiveBlack-II} galaxies, we only considered those with $M_{\mathrm{stars}} > 5 \times 10^8$ M$_{\odot}$, matching the mass range used in the \citet{elbaz07} fit.  Fig. \ref{fig:smsfr} shows a clear offset between the observed relation and the \emph{MassiveBlack-II} galaxies, which exists for all techniques, with an RMS deviation of order 1.5 dex.  The best-fitting slopes match the M12 systems almost identically.  For both simulation sets then, there is no distinct technique that shows superior agreement for this scaling relation.

The relation derived by \citet{elbaz07} was done with blue galaxies, defined by colour index $U - g < 1.45$.  However, an equivalent cut for the \emph{MassiveBlack-II} galaxies does not produce a conducive comparison, as at low redshift, they are all red; using theoretical spectra for each \emph{MassiveBlack-II} subhalo \citep[for how these were derived, see][]{khandai14}, we found only $\sim$2 per cent of galaxies to have $U - g < 1.45$.  The fact that all the galaxies are red is consistent with their low (specific and standard) star formation rates.  Hydrodynamic simulations have been known to produce galaxies with low star formation rates and high stellar masses at low redshift; for example, see \citet{keres09b}, who show that by normalising simulated galaxy masses to match the observed stellar mass function, one can also match the observed stellar mass--specific star formation rate relation (both their simulation and \emph{MassiveBlack-II} produce too many low-mass and high-mass systems).

\subsubsection{Kennicutt-Schmidt relation}

The Kennicutt-Schmidt relation is a well-known scaling relation, where the average star formation rate surface density goes approximately as a power law of the average gas surface density, i.e. $\Sigma_{\mathrm{SFR}} \approxprop \Sigma_{\mathrm{gas}}^{\ \ \ n}$. Specifically, \citet{kennicutt98} formulated an analytic relationship:
\begin{equation}
\label{eq:ks}
\frac{\Sigma_{\mathrm{SFR}}}{\mathrm{M}_{\odot}\ \mathrm{yr}^{-1}\ \mathrm{pc}^{-2}} = 
\frac{2.5 \pm 0.7}{10^{10}} 
\left( \frac{\Sigma_{\mathrm{gas}}}{\mathrm{M}_{\odot}\ \mathrm{pc}^{-2}} \right)^{1.40 \pm 0.15}\ .
\end{equation}
\noindent  We note that Equation \ref{eq:ks} was derived taking galaxies to end at their optical limit.  Nevertheless, we find it informative to assess whether using alternative definitions for the end of a galaxy affect how well simulations match the observed relation or whether they produce an alternative equivalent relationship.

We have combined measurements of star formation rate, gas mass, and aperture radius from each technique in Table \ref{tab:apertures} to produce $\Sigma_{\mathrm{gas}}$ and $\Sigma_{\mathrm{SFR}}$ values, as plotted in Fig. \ref{fig:ks}.  For those techniques plotted, subhalo finders were included where possible (denoted in Table \ref{tab:apertures}).  The surface area in each case was taken to be $\pi r_{\mathrm{aper}}^{\ \ \ \ 2}$.  We only considered cold gas to contribute toward $\Sigma_{\mathrm{gas}}$, even for the direct aperture techniques.  For clarity, we use dots to only show the $z=0$ case for the M12 galaxies in Fig. \ref{fig:ks}, but data from all $z < 2$ snapshots were combined when fitting the straight-line relationships. Again, contours represent the \emph{MassiveBlack-II} galaxies.

\begin{figure}[t!]
\centering
\includegraphics[width=\textwidth]{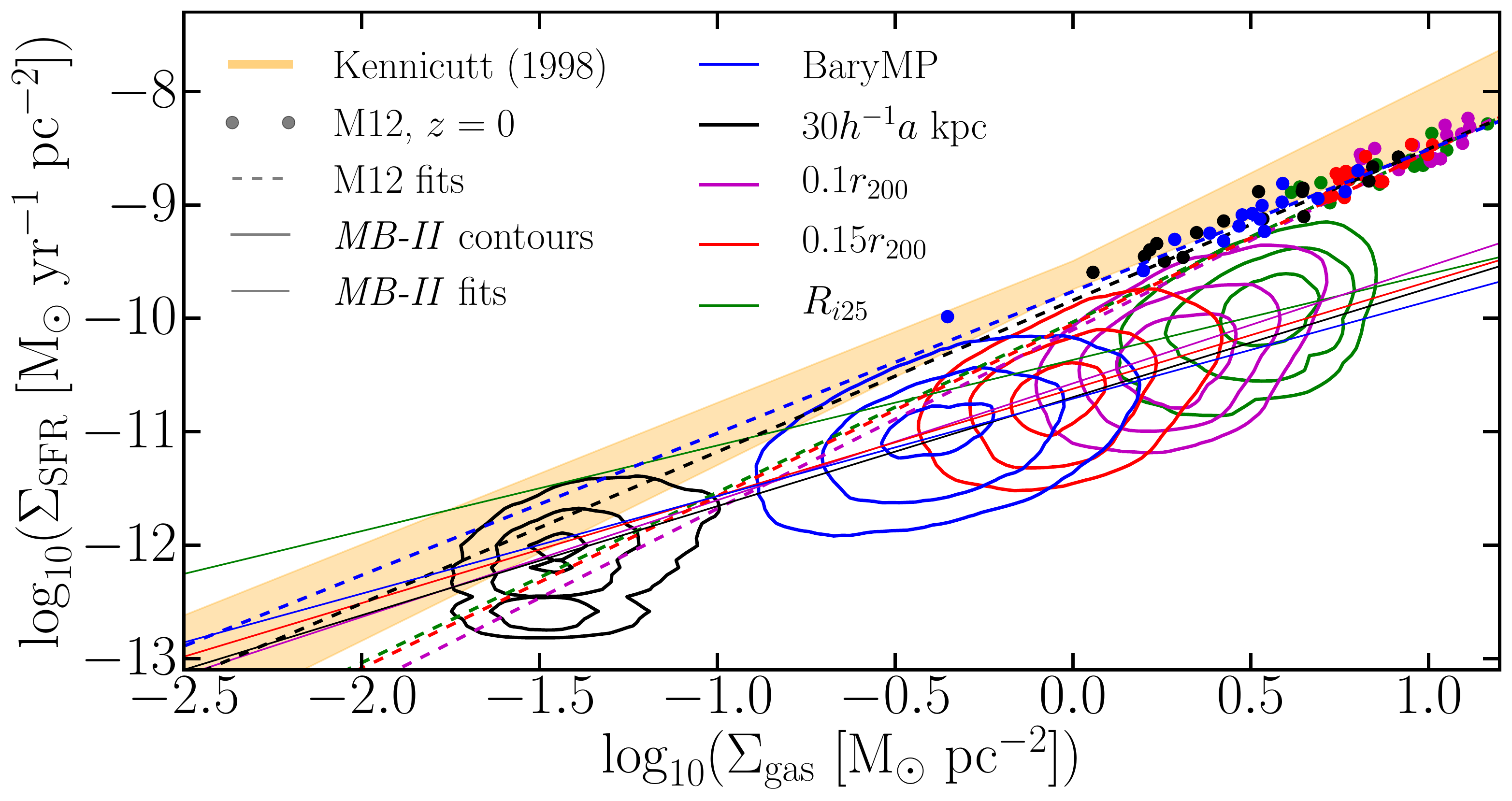}
\caption{Kennicutt-Schmidt relation for each aperture technique (with subhalo finders applied where listed in Table \ref{tab:apertures}). Dots represent the M12 galaxies at $z=0$. \emph{MassiveBlack-II} galaxies are represented with contours.  The orange shaded region follows Equation \ref{eq:ks}.  The best-fitting lines to each dataset (which included all snapshots of the M12 galaxies where $z<2$) indicate technique-independence for falling within the shaded region, but that the derived slope can vary with technique choice.}
\label{fig:ks}
\end{figure}

For the M12 galaxies, we found the best-fitting power-law slope, $n$, to be between 1.17 and 1.58 for the techniques listed in Table \ref{tab:apertures}.  The slope of the relation for the M12 galaxies varies negligibly for a given aperture applied with and without the additional use of \textsc{ahf}.  The \emph{MassiveBlack-II} systems instead give values of $n$ between 0.86 and 1.02.  The M12 systems show a much smaller level of scatter, with RMS deviations $\lesssim$0.20 dex, while the \emph{MassiveBlack-II} systems instead show an RMS of order 0.5 dex for each technique. While the best-fitting slopes in many cases do not fall in the quoted range of \citet{kennicutt98}, we found the RMS for a forced fit of $n = 1.4$ to be scarcely different for when $n$ was fitted in each case ($<$10 per cent increase), with the M12 galaxies also showing general agreement with Equation \ref{eq:ks} for each technique.

Evident from Fig. \ref{fig:ks}, the average \emph{MassiveBlack-II} galaxy has a low $\Sigma_{\mathrm{gas}}$ compared to the M12 galaxies (and observed galaxies, for that matter).  This is consistent with the galaxies' low concentration.  However, spreading gas and star formation over a larger area will not maintain agreement with the Kennicutt-Schmidt relation unless the overall star formation rate decreases too (as $n > 1$).  As discussed in Section \ref{ssec:sfr}, these galaxies do have low star formation rates.  As such, their offset from the observed Kennicutt-Schmidt relation is small (and less than from the observed stellar mass--star formation rate relation).

\section{Discussion}
\label{sec:discuss}
\subsection{Sizes of the simulated galaxies}

To claim our results impact the post-processing of hydrodynamic simulations in general, one should address whether our sample of simulated galaxies is representative. If the `full stellar mass' is calculated using all the star particles within a (sub)halo (no substructure), then, for stellar half-mass radii of the M12 galaxies, we find $0.021 < r_{1/2} / r_{200} < 0.041$ at $z=0$. These values fall within the observed range of \citet[][our Equation \ref{eq:rhalf}]{kravtsov13}. While this means the sizes of the galaxies are physically reasonable \citep[for more on the sizes of the galaxies, see][]{martig12}, it does suggest our sample is biased by lower-concentration galaxies compared to an observed sample.  Consistent with the notion that the \emph{MassiveBlack-II} galaxies are even less concentrated, $r_{1/2} / r_{200}$ values for \emph{MassiveBlack-II} are larger than M12 on average, with most larger than observed galaxies.

It should be noted that the differences in measured properties from the techniques that invoke an aperture of a fixed fraction of the virial radius versus the other techniques would be less extreme in a sample of more-concentrated galaxies.  However, if a given technique fails to capture a simulated galaxy in the less-concentrated instances, it fails to be generally applicable.   After all, if a simulation produces all low-concentration galaxies, the properties of those galaxies will likely not be discarded, as they still have scientific merit.  Their low concentration may not be noted in the published results either.  The fact that the concentration of simulated galaxies changes for different redshifts, formation histories, and the design of the simulation itself, reinforces the notion that a fixed fraction of the virial radius can not define the general extent of galaxies. Ultimately, the M12 and \emph{MassiveBlack-II} galaxies provide as fair a tool as any to compare the aperture techniques.

\subsection{Which method should I choose?}

As outlined in Section \ref{sec:techs}, we suggest that if one wants to study the evolution of a galaxy in a simulated (sub)halo, particles that are attached to any form of substructure, or that are diffusely spread throughout the (sub)halo, should not contribute.  Even for single-halo simulations, we therefore encourage using a subhalo finder as a first step, and subsequently the exclusion of hot, low-density gas.  We further encourage an aperture technique, not only to discern galaxies from the rest of their parent (sub)haloes, but also because no subhalo finder is perfect at locating all substructure to be removed (as seen in Fig. \ref{fig:images2}).

It is clear some of the aperture techniques assessed in this paper failed to encompass a reasonable fraction of the galaxy. As already discussed, $0.1r_{200}$ and $0.15r_{200}$ poorly define the edge of a galaxy, due to the variation in the galaxy-to-virial radius fraction for (sub)haloes. Despite these techniques' popularity, we advise against them.

While a technique such as $R_{i25}$ is fair for comparison to some observations, the additional level of stellar-population modelling required and the technique's common trait of cutting out a legitimate portion of the galaxy make it arguably less than ideal for an equally footed technique across all simulations. Because stellar mass and brightness track each other to first order, a technique that invokes an absolute cut-off in stellar surface density would prove to similar effect, with simpler implementation, and would not force observational constraints which impede measuring a galaxy's \emph{true} integrated properties.  

We suggest the most sensible approach for a technique to measure any and all galaxy properties self-consistently is one that includes an aperture whose radius is directly determined by the baryonic content within each respective (sub)halo.  While the suggestion of Figs. \ref{fig:smrel}, \ref{fig:gmrel}, and \ref{fig:mbiirel} is that the additional application of such an aperture, BaryMP, would alter values of gas and stellar mass by only a few per cent on average, there are a sufficient number of instances where a greater difference occurs for this to have a significant effect.  Gas accretion and ejection rates also show a large difference with the application of the aperture versus none.  BaryMP comes with an additional advantage of being unaffected by straggling satellites not identified by the subhalo finder (see Appendix \ref{app:algorithm}).

It is totally within reason that the scope of one's study might not mean that a highly detailed analysis on which particles are part of the main galaxy of a (sub)halo is necessary. If, for example, one desired to measure just the integrated stellar mass of a Milky-Way-like simulated galaxy, then any technique used in this paper would be appropriate to return an answer approximately with a 20-per-cent uncertainty. In such a case, to minimize computation and effort, a direct aperture technique would be sensible.

\section{Conclusion}
\label{sec:conc}
When studying the gross properties of galaxies from hydrodynamic simulations, the first step is to determine which particles/cells should contribute to those properties. Throughout the published literature, a number of different techniques has been used for this step. Using a subsample of the high-resolution simulations of \citet{martig12} and a selection of galaxies from the \emph{MassiveBlack-II} simulation, the integrated properties of galaxies were measured using a range of different techniques, including several from the literature. These techniques include the use of a subhalo finder, temperature constraint for gas, and/or spherical apertures, the latter defining where galaxies end.

Comparison of the two most popular techniques in the literature [an aperture of $0.1r_{200}$ versus the entire (sub)halo with substructure removed] shows differences in the average Milky-Way-mass system of order 30 per cent for stellar mass, a factor of 3 for gas mass, and 40 per cent for star formation rate.  Gas accretion and ejection rates are the most susceptible properties to technique choice, with variations of an order of magnitude not unlikely.  The choice of technique is hence key to the interpretation of simulation results.  Our Table \ref{tab:dists} further details these differences and the standard deviation of the distributions that go alongside them.

For a more general population of galaxies (stellar and gas mass each between $10^8$ and $10^{12}$ M$_{\odot}$), gas mass and especially star formation rate measurements exhibit less susceptibility to technique, assuming only cold gas is considered, but can still show significant differences for systems where the gaseous and stellar centres of mass are offset.  Stellar mass, on the other hand, shows a similar level of variation to the Milky-Way-mass population, with the exception of the observationally motivated $R_{i25}$ aperture.

Among the techniques we compared, in attempt to separate a galaxy from its diffuse baryonic surroundings in a physically motivated manner, we defined a new aperture technique, whose radius is determined from the cumulative (cold) baryonic mass profiles of (sub)haloes.  While, on average, each integrated property measurement was only a few per cent less than that of the full (sub)halo, the distributive nature of these differences highlights the importance of applying such an aperture to obtain universally fair values.  The average gas mass and star formation rate measurements were also significantly less than the full (sub)halo for the Milky-Way-mass \emph{MassiveBlack-II} systems (\frenchspacing{cf. Table} \ref{tab:dists}).

While the choice of technique for the measurement for individual integrated properties is important, we have shown that scaling relations do not rely heavily on technique to match observations.  One can, however, derive a difference in the slope of the Kennicutt-Schmidt relation of order 25 per cent using a single simulation set based on the various measurement techniques we have assessed.

For the sake of clarity, we encourage all authors to be explicit in how they have defined galaxies within their simulations, but more importantly to discuss and justify their choice of technique.  Our analysis sheds light on the uncertainties associated with comparing results of simulations that use different techniques to measure galaxy properties.  While there may not be a perfect technique, we have shown that apertures defined by a constant fraction of the virial radius do not succeed on a general basis, and better alternatives exist.

While the question `Where do galaxies end?' is difficult to answer, it is an important question to ask if we are to have a common understanding of the properties of galaxies, both from simulation and observation perspectives, which ultimately impacts our understanding of the underlying astrophysics that generates these properties.

\section*{Acknowledgements}

We thank Tiziana Di Matteo and Nishikanta Khandai for access to the \emph{MassiveBlack-II} simulation and responding so positively to its use in this project.  We thank the anonymous referee for suggestions and queries that ultimately led to the improvement of this paper.  ARHS would also like to thank Chiara Tonini and Chris Brook for conversations related to this work.  Analysis of the M12 simulations was primarily performed on the gSTAR supercomputer at Swinburne University of Technology.  The \emph{MassiveBlack-II} analysis was performed on the Ferrari cluster at the McWilliams Center for Cosmology, using the on-site particle data.  DJC acknowledges receipt of a QEII Fellowship by the Australian Research Council (DP1095506).

\appendix

\section{The B\lowercase{ary}MP algorithm}
\label{app:algorithm}

We know the gradient of the baryonic mass profiles must be steeper than $m$ (Equation \ref{eq:barymp}) for $r \ll r_{200}$, as baryons are more concentrated toward the centre of (sub)haloes (\frenchspacing{i.e. in} galaxies).  This places an upper limit on $m$ of 1.  Also, because the profiles are cumulative, $m \geq 0$.  As such, our algorithm begins by locating the point on each profile where the gradient is 1, following its initial climb.  Because $m<1$, this provides a lower limit for the aperture radius.  The profile from this point onward is then extracted and a straight line fitted.  Parts of the profile where the cumulative baryonic mass fraction exceeds a separation of $\epsilon$ from the fitted line are then removed and the line refitted.  This is done continuously until a converged result is found.  The lowest of the remaining points of the profile is then taken to be the radius of the aperture.

\begin{figure}[t!]
\centering
\includegraphics[width=\textwidth]{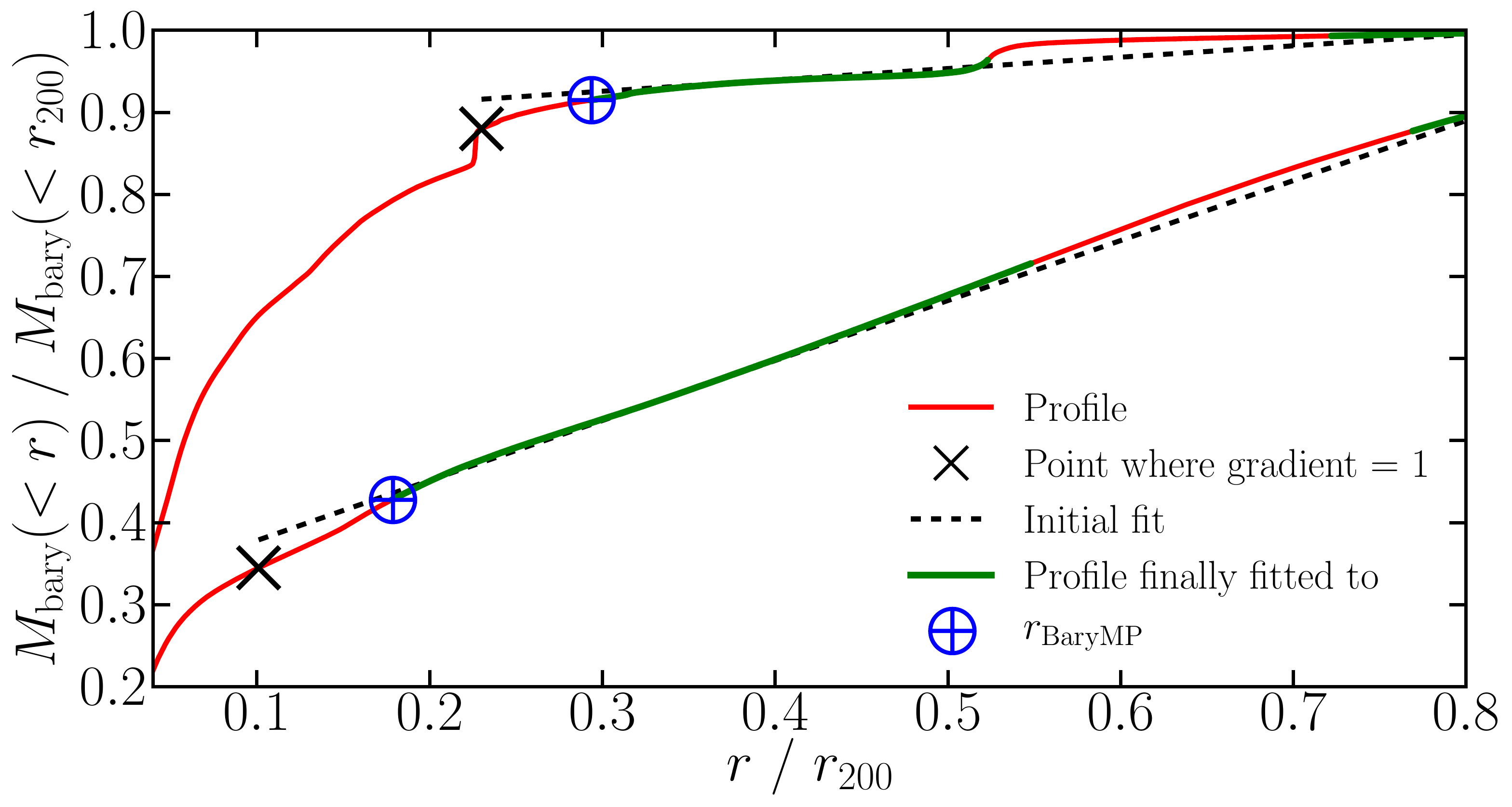}
\caption{Illustration of the BaryMP algorithm for two example M12 haloes.  Red, solid curves show the baryonic mass profiles of the haloes (with \textsc{ahf}-identified substructure and hot gas stripped).  Black crosses mark where the profiles have a gradient of 1, while black, dashed lines show the initial linear fit from these points onward.  After several iterations of removing parts of the profile separated from the linear fit by $\epsilon=0.01$ and refitting the line, the final parts of the profile fitted are given in green.  The aperture radius is taken at the lowest-radius point on the green parts of each profile, emphasized by the blue plus-filled circles.  The effect of bumps in profiles caused by unidentified satellites are shown to be minimized by the algorithm.  The final linear fits for these examples almost overlay the initial fits exactly (but are not shown).}
\label{fig:fitting}
\end{figure}

For the results presented in this paper, we applied the algorithm with $\epsilon = 0.01$.  A sufficiently small $\epsilon$ is needed to return an aperture radius above the lower limit, while a value too small would exclude the majority of the profile from the final fit.  An appropriately chosen $\epsilon$ also comes with the advantage of dealing with late rises in profiles caused by satellites and other substructure (no subhalo finder is perfect at locating all satellites, as was the case on several occasions for these simulations, \frenchspacing{e.g. as} presented in Fig. \ref{fig:images2}).  The top profile in Fig. \ref{fig:fitting} displays the success here with $\epsilon = 0.01$.  Should any substructure lie beyond the lower limit of the aperture, the final radius will be found internal to that substructure, hence eliminating its contribution to the (sub)halo.  Should sharp rises in the profiles occur internal to the lower limit, we determine the responsible substructure to be sufficiently close to / merged with the central galaxy to be considered the same system (also seen in the top profile of Fig. \ref{fig:fitting}).  We examine the dependence of BaryMP on $\epsilon$ in Appendix \ref{app:epsilon}.

Below we provide Python code for our algorithm to calculate the radius of a galaxy from the BaryMP method.  The inputs \texttt{x} and \texttt{y} represent one-dimensional arrays for corresponding values of $r/r_{200}$ and $M_{\mathrm{bary}}(<r) / M_{\mathrm{bary}}(<r_{200})$, respectively.  \texttt{eps} is the value of $\epsilon$, while \texttt{r\_bmp} is $r_{\mathrm{BaryMP}} / r_{200}$.

\begin{lstlisting}
import numpy as np
def BaryMP(x,y,eps=0.01):
	dydx = np.diff(y)/np.diff(x)
	
    maxarg = np.argwhere(dydx==np.max(dydx))[0][0] # Find where the gradient peaks
    xind = np.argwhere(dydx[maxarg:]<=1)[0][0] + maxarg # The index where the gradient reaches 1
	
    x2fit_new, y2fit_new = x[xind:], y[xind:]
    x2fit, y2fit = np.array([]), np.array([])
	
    while len(y2fit)!=len(y2fit_new):
        x2fit = np.array(x2fit_new)
        y2fit = np.array(y2fit_new)
        p = np.polyfit(x2fit, y2fit, 1)
        yfit = p[0]*x2fit + p[1]
        sep = abs(yfit-y2fit)
        sepf = (chi<eps)
        x2fit_new = x2fit[sepf]
        y2fit_new = y2fit[sepf]
    r_bmp = x2fit[0]
    return r_bmp
\end{lstlisting}

\section{The effect of $\mathlarger{\mathlarger{\epsilon}}$ on the B\lowercase{ary}MP Technique}
\label{app:epsilon}

While we qualitatively described that choosing $\epsilon = 0.01$ works reasonably well in Appendix \ref{app:algorithm}, there was no specific motivation behind this exact value.  As such, we briefly assess the variation in BaryMP results induced from different choices of $\epsilon$.  To do this, we recalculated the BaryMP radii for each snapshot of the M12 simulations where $z<2$ for 198 uniformly spaced values of $\epsilon$ between $10^{-3}$ and $10^{-1}$, recording the (cold) baryonic content within each radius and the fraction of the profile (beyond the lower limit, where the gradient is 1) used in the final linear fit.  We present the average of each of these values as a function of $\epsilon$ in Fig. \ref{fig:epsilon}.

\begin{figure}[t!]
\includegraphics[width=\textwidth]{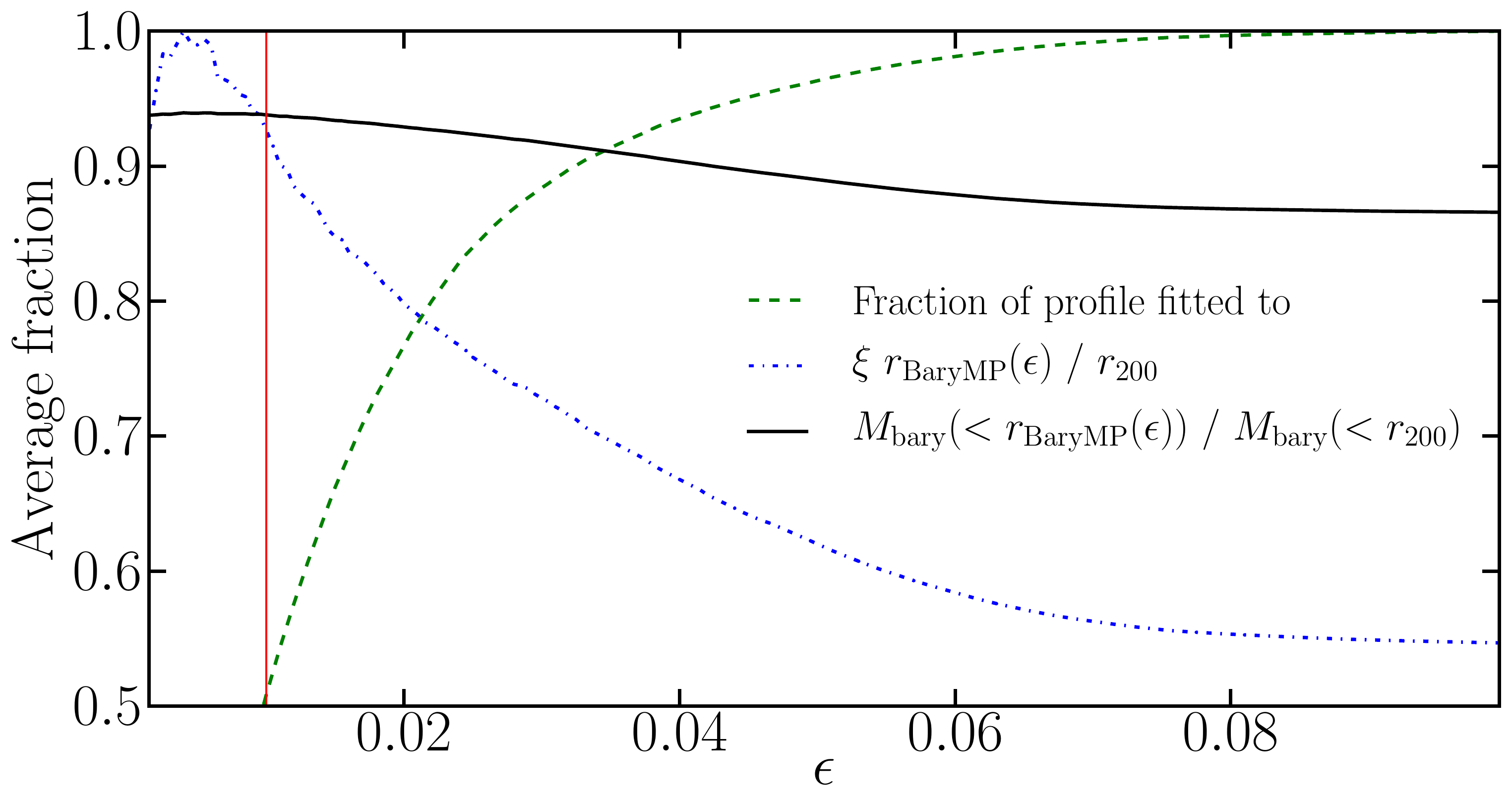}
\caption{Average variation in results from the BaryMP technique for the M12 simulations as a function of the profile-fit separation threshold, $\epsilon$.  The green, dashed curve gives the fraction of the profile beyond the lower limit used in final linear fit.  The blue, dot-dashed curve gives the average aperture radius (as a fraction of the virial radius) inferred from that fit.  Note that these values have been normalised to fit within the axes ($\xi = 2.745$).  The black, solid curve indicates the (cold) baryonic content within those apertures.  The red, vertical line shows our choice of $\epsilon=0.01$.}
\label{fig:epsilon}
\end{figure}

On average, for $\epsilon \gtrsim 0.1$, the linear fit is converged on its first attempt.  That is, the entirety of the profile beyond the lower limit deviates by $<$0.1 in the $y$-direction from the initial fit.  This places an obvious upper limit on an appropriate value of $\epsilon$. The BaryMP technique's success lies in its ability to ignore both early parts of the profile with varying gradient and straggling satellites that cause bumps later in the profile.  In other words, it works well if a fraction, but not too large of a fraction, of the profile is removed for the final linear fit.  Fig. \ref{fig:epsilon} therefore suggests it would be appropriate if $0.01 \lesssim \epsilon \lesssim 0.04$ (those values corresponding to average profile fractions of 0.5 and 0.93, respectively).  Within this range, the average $r_{\mathrm{BaryMP}}/r_{200}$ value clearly varies.  Despite this, however, the variation in the average enclosed baryonic content is small.

With Fig. \ref{fig:epsilon2}, we specifically assess how the BaryMP aperture radius and enclosed baryonic content would change relative to our currently presented results if we picked a different $\epsilon$ between 0.01 and 0.04.  A choice of $\epsilon = 0.04$ would decrease the average BaryMP radius by 24 per cent, with individual cases potentially having their radii halved.  These rarer cases occur where an especially large radius was found for $\epsilon=0.01$ (a prominent bump in the blue curve in Fig. \ref{fig:mbiirel}a and subtler bumps in Figs. \ref{fig:radrel}b and \ref{fig:radrel}c all show small populations of galaxies with $r_{\mathrm{BaryMP}} \gtrsim 0.6 r_{200}$).  These anomalies crop up at various values of $\epsilon$, but do not greatly affect the values of integrated properties. As shown by Fig. \ref{fig:epsilon2}, the average enclosed baryonic mass decreases by only 4 per cent for $\epsilon=0.04$, where decreases of more than 6 per cent happen for less than 16 per cent of the M12 simulation snapshots.  

The noticeable dependence of $r_{\mathrm{BaryMP}}$ on $\epsilon$ emphasizes the difficulty of defining or locating where a galaxy ends.   The lack of dependence on $\epsilon$ for the baryonic mass suggests our choice of $\epsilon$ should not affect our conclusions surrounding the integrated properties of galaxies, and shows that the BaryMP method provides a relatively consistent means of measuring them.

\begin{figure}[t!]
\includegraphics[width=\textwidth]{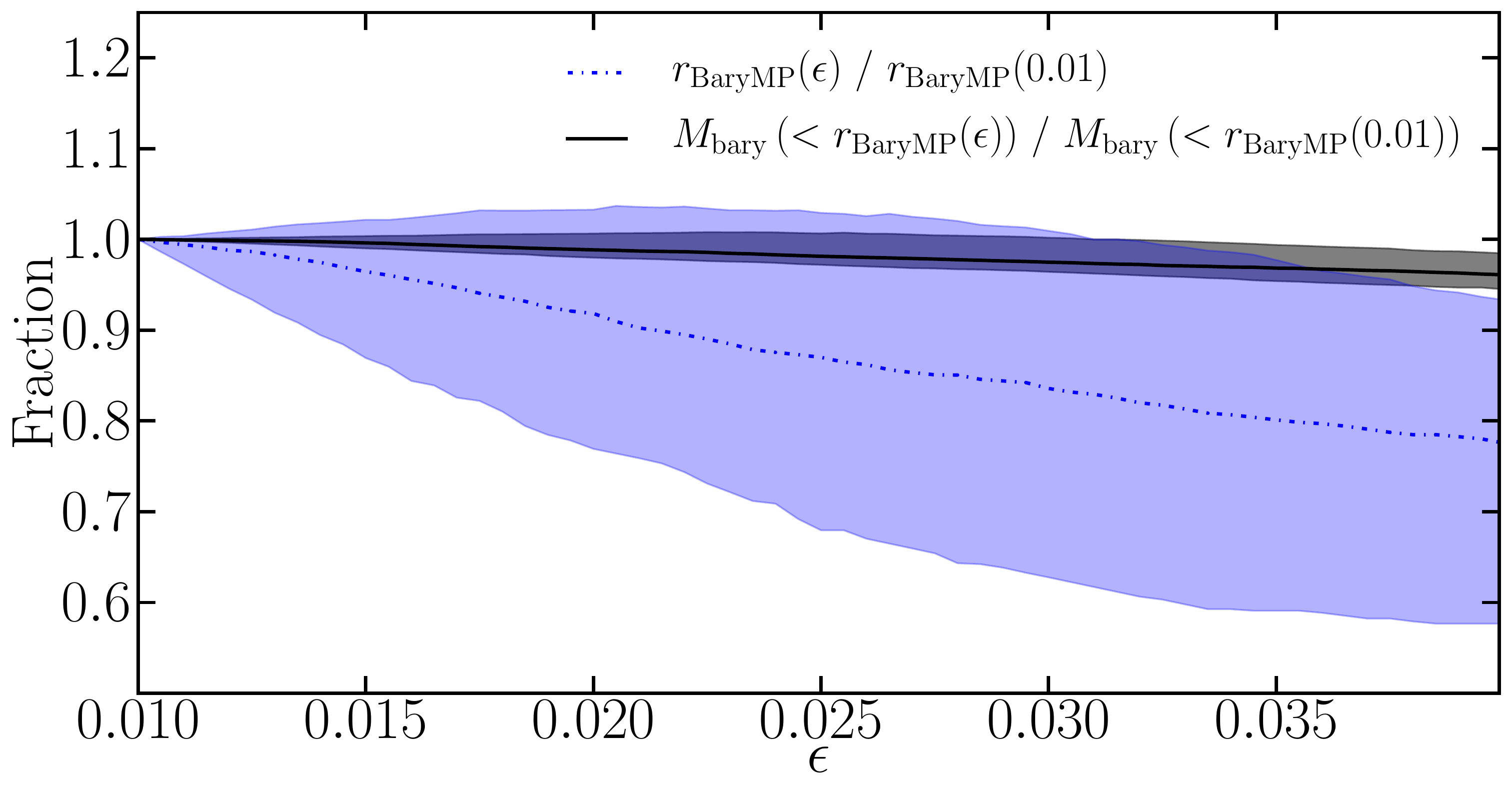}
\caption{BaryMP aperture radius and enclosed (cold) baryonic mass as a function of $\epsilon$, normalised to the choice of $\epsilon=0.01$. The blue, dot-dashed curve gives the median aperture radius over all the M12 simulation snapshots for $z<2$, with the blue shaded region covering the central 68 per cent of the values.  Similarly, the black curve and shaded region show the median and 68-per-cent spread for the baryonic mass, respectively.}
\label{fig:epsilon2}
\end{figure}

\end{document}